\documentclass{cernrep}

\usepackage{amsmath,amssymb}
\usepackage{graphicx}
\usepackage{booktabs}
\usepackage{hyperref}

\usepackage[T1]{fontenc}
\usepackage{graphicx}
\usepackage{amsmath}
\usepackage{slashed}
\usepackage{braket}
\usepackage[export]{adjustbox}
\usepackage{enumitem}
\usepackage{placeins} 
\usepackage[normalem]{ulem}
\usepackage{xspace}
\usepackage{bbm}
\usepackage{xcolor}
\usepackage[
    top=20mm,
    bottom=25mm,
    left=20mm,
    right=20mm
 ]{geometry}
\usepackage{tikz}
\usepackage{eso-pic}

\newcommand{\df}{\mathrm{d}}

\newcommand{\GeV}{\,\mathrm{GeV}}

\newcommand{\cO}{\mathcal{O}}

\newcommand{\as}{\alpha_s}

\newcommand{\lqcd}{\Lambda_\mathrm{QCD}}

\newcommand{\refcite}[1]{ref.~\cite{#1}}
\newcommand{\refscite}[1]{refs.~\cite{#1}}
\renewcommand{\Eq}[1]{Eq.~\eqref{eq:#1}}

\newcommand{\fig}[1]{figure~\ref{fig:#1}}

\def\be{\begin{equation}}
\def\ee{\end{equation}}
\def\beq{\begin{equation}}
\def\eeq{\end{equation}}

\newcommand{\alphas}{\ensuremath{\alpha_s}\xspace}

\begin{document}
\AddToShipoutPictureFG*{%
  \begin{tikzpicture}[remember picture,overlay]
    \node[anchor=north east]
      at ([xshift=-2cm,yshift=-1cm]current page.north east)
      {CERN-TH-2026-211, MIT-CTP 6100, MPP-2026-147, UWThPh  2026-9};
  \end{tikzpicture}%
}
\title{A Task Force on Strong Coupling Determinations from Event Shapes}
\author{\large {\bf Editors}: A.~Badea$^{1}$, L.~Buonocore$^{2}$, and G.~Vita$^{2}$\\[.2cm]
{\bf Contributors}: S. Alioli$^{3,4}$, Z. Capatti$^{5}$, Y. Chen$^{6}$, D. d'Enterria$^{7}$, T. Gehrmann$^{8}$, A. H. Hoang$^{9}$, A. Huss$^{2}$, S. Jaskiewicz$^{5}$, A. Karlberg$^{10}$, Y. Lee$^{11}$, M. L. Mangano$^{2}$, V. Mateu$^{12}$, J. Miao$^{13}$, P. Monni$^{2}$, \\[.5mm] 
P. Nason$^{4}$, L. Rottoli$^{3,4}$, I. W. Stewart$^{14}$, P. Torrielli$^{15}$, and G. Zanderighi$^{10,16}$
\vspace*{.5cm}}
\institute{
$^{1}$ University of Chicago, Enrico Fermi Institute, 60637 IL, USA,\\
$^{2}$ CERN, TH Department, CH-1211 Geneva 23, Switzerland,\\
$^{3}$ Dipartimento di Fisica G. Occhialini, Università degli Studi di Milano-Bicocca,\\
$^{4}$ INFN, Sezione di Milano-Bicocca, Piazza della Scienza 3, 20126 Milano, Italy,\\
$^{5}$ University of Bern, Sidlerstrasse 5, 3012 Bern, Switzerland,\\
$^{6}$ Vanderbilt University, Nashville, TN 37215, USA,\\
$^{7}$ CERN, EP Department, CH-1211 Geneva 23, Switzerland,\\
$^{8}$ Physik-Institut, Universität Zürich, CH-8057 Zürich, Switzerland,\\
$^{9}$ University of Vienna, Faculty of Physics, Boltzmanngasse 5, A-1090 Wien, Austria\\
$^{10}$ Max Planck Institute for Physics, Boltzmannstr. 8, 85748 Garching, Germany,\\
$^{11}$ MIT, Laboratory for Nuclear Science, Cambridge, MA 02139, USA,\\
$^{12}$ Fundamental Physics Department and IUFFyM, University of Salamanca, E-37008 Salamanca, Spain,\\
$^{13}$ Dipartimento di Fisica A. Pontremoli, Università degli Studi di Milano, and INFN, Sezione di Milano, 20133 Milano, Italy,\\
$^{14}$ MIT Center for Theoretical Physics -- A Leinweber Institute, Cambridge, MA 02139, USA,\\
$^{15}$ Dipartimento di Fisica, Università degli Studi di Torino, and INFN, Sezione di Torino, Via P. Giuria 1, 10125 Torino, Italy,\\
$^{16}$ Physik Department T31, Technische Universität München, James-Franck-Straße 1, D-85748 Garching, Germany
}
\begin{abstract} %
The strong coupling constant $\alpha_s$ is a fundamental parameter of the Standard Model. 
Its precise determination is essential for accurately predicting, studying, and understanding processes at the Large Hadron Collider and future experiments such as the Future Circular Collider.
Event shape and correlator observables measured at electron-positron colliders provide one of the cleanest environments for extracting \alphas, thanks to their sensitivity to $\alphas$ and the availability of high-precision data from the Large Electron-Positron Collider.
More broadly, such observables provide an ideal setting to develop and test our understanding of the perturbative and non-perturbative elements of Quantum Chromodynamics, which will underpin the field’s precision and discovery frontiers for decades to come.
Despite these advances, significant discrepancies persist between different determinations of \alphas from event shapes, both in the extracted central values and estimated uncertainties.
This document motivates the establishment of a dedicated Task Force to coordinate a community-wide effort addressing these open questions.
We report on the first two-day meeting held at CERN in November 2025, summarizing the scientific discussion and documenting the experimental analyses identified as priorities during the meeting, as well as the concrete list of tasks to be carried out by the theory community in preparation for future meetings.
\end{abstract}
\maketitle
\newpage
\section*{Executive summary}
\label{sec:executive}

This document motivates the establishment of a new Task Force on Strong Coupling Determinations from Event Shapes, a coordinated platform bringing together theorists and experimentalists working on precision event shape observables. 
Its scope is to provide a framework in which different methodologies can be compared on transparent and technically well-defined grounds, developing common benchmarks and validation tests for sources of theoretical and experimental uncertainties in collider determinations of $\alpha_s$. 

A dedicated effort of this kind is needed because significant progress on these issues requires sustained interaction, detailed cross-checks, and a coordinated programme of follow-up studies on narrow technical points that cannot be accommodated within the format of standard conference and workshop discussions.

As an initial step, the task force held a two-day kick-off meeting at CERN on November 27--28, 2025, focused on four scientific themes in thrust-based determinations of $\alpha_s$~\cite{indico}. 
These topics were chosen as a concrete starting point for testing the working format, not as a boundary on the task-force mandate. Several important points and follow-up priorities emerged from the kick-off meeting. 
Additional observables and themes will be addressed in future meetings according to the needs and requests of the community.
 
 First, a productive discussion took place on 3-jet power corrections. This made it clear that a more fundamental issue needs to be resolved first: there is no consensus on the criteria to identify the validity of the dijet approximation even at the perturbative level and therefore to establish when including resummation is necessary, appropriate, or potentially detrimental to the accuracy of a prediction. 
Given the central role of resummed predictions in precision collider phenomenology, well beyond $\alpha_s$ determinations from thrust, the development of clear quantitative tests for assessing the domain of validity of resummed predictions in a way that can be applied consistently and universally across observables and processes was identified as an urgent priority for the task force. 
Two concrete follow-up exercises were identified during the workshop. 

Second, the usefulness and the limits of the field-theoretical treatment of hadron-mass effects were discussed.
In a single-observable fit genuinely restricted to the dijet limit, the leading hadron-mass dependence is absorbed into a single non-perturbative parameter.
A key question is whether this assumption is sufficiently justified in the fit window employed.
To address this, dijet fits will be performed in both the $p$- and $E$-schemes. A significant shift in $\alpha_s$ would signal a breakdown of the above assumption.
Another aspect is that in multi-observable fits, or in fits away from the dijet region, universality assumptions do not hold, and imposing a single common non-perturbative parameter can bias the extracted $\alpha_s$.
A concrete follow-up is therefore to repeat joint fits with multiple non-perturbative parameters, assigned according to the relevant universality classes and fit regions.
Dedicated experimental measurements in both the $p$- and $E$-schemes are an essential part of this programme, since they would make the scheme-comparison tests substantially more robust and reduce the reliance on Monte Carlo based transformations between schemes.

Third, the $e^+e^-$ alliance presented preliminary ALEPH and DELPHI reanalyses that indicate shifts with respect to the original LEP-era thrust results whose size and location could have significant impacts on modern determinations of $\alpha_s$, comparable to, or larger than, theoretical systematics that are currently the subject of detailed debate. This makes continued LEP reanalyses, improved documentation of experimental systematics, and the provision of correlations across observables central experimental priorities for the task force. 

Fourth, a large difference between predictions performed in two formally equivalent resummation spaces in the thrust fit region had been reported in one analysis. 
However, this was not reproduced by two independent groups who presented their analyses at the workshop. 
In the regions usually employed for \alphas fits, they both found significantly smaller differences at the highest perturbative order and a common qualitative picture in which the resummation-space ambiguity decreases with increasing perturbative accuracy. 
The discussion broadly supported the view that neither space should be treated as intrinsically preferred at finite logarithmic accuracy, and that any residual ambiguity should be encompassed by a reliable perturbative uncertainty estimate. A follow-up exercise was identified to understand why the large discrepancies reported in one analysis were not reproduced by the other studies.

These follow-up studies define the immediate technical programme of the task force.
Their purpose is not to impose a single analysis framework, but to provide reference tests, shared benchmarks, and clearly documented assumptions for future high-precision determinations of $\alpha_s$. 
In the longer term, the task force aims to produce a comprehensive assessment analogous to community-based efforts for lattice QCD determinations.

\section{Introduction and Motivation}

The strong coupling constant, $\alpha_s$, is one of the fundamental parameters of the Standard Model. Its precise determination is essential for quantitative tests of Quantum Chromodynamics (QCD), interpretation of precision electroweak measurements, and searches for new physics. 
The value of $\alpha_s$ governs the dynamics of the strong interaction, influences the internal structure of the proton, and enters parametrically into theoretical predictions for processes at the LHC and future high-energy colliders. Over the past decades, competitive determinations of $\alpha_s$ have been obtained from high-precision measurements of event-shape observables in the clean environment provided by $e^+e^-$ collisions.
\begin{figure}[ht]
\centering
\includegraphics[width=0.3\textwidth]{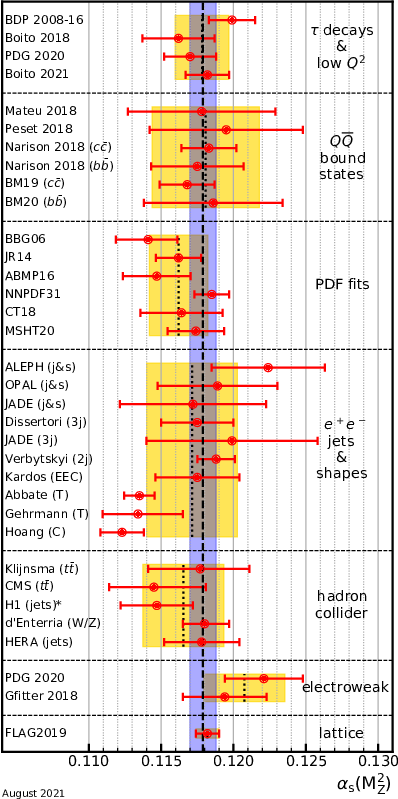} \hspace{2.5cm}
\includegraphics[width=0.28\textwidth]{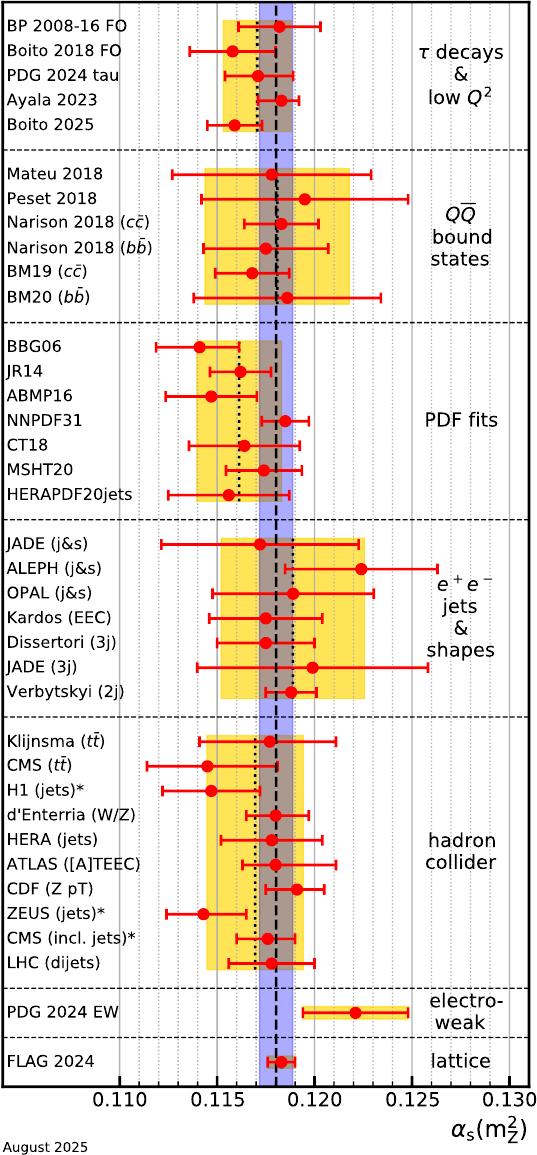}
\caption{PDG summaries of \(\alpha_s(m_Z)\) determinations in 2021~\cite{PDG2020} (left) and in the latest 2025 update~\cite{ParticleDataGroup:2024cfk} (right). The 2021 compilation includes extractions based on higher-order resummation matched to fixed order with analytic power-correction frameworks, which have been excluded from the PDG world average since the 2023 revision \cite{PDG2022}.}
\label{fig:PDG}
\end{figure}

The success of this programme has been enabled by remarkable progress in the theoretical description of event-shape observables.
The current state of the art includes fixed-order predictions up to third order in the strong coupling, $\mathcal{O}(\alpha_s^3)$~\cite{Gehrmann-DeRidder:2014hxk,DelDuca:2016ily,Weinzierl:2009ms,NNLOJET:2025rno}, as well as leading-power resummations reaching next-to-next-to-next-to-next-to-leading logarithmic (N$^4$LL) accuracy~\cite{Becher:2008cf,Baikov:2009bg,Gehrmann:2010ue,Abbate:2010xh,Monni:2011gb,Abbate:2012jh,Gehrmann:2012sc,Mateu:2013gya,Hoang:2014wka,Hoang:2015hka,Li:2016ctv,Monni:2016ktx,Baikov:2016tgj,Herzog:2017ohr,Bizon:2017rah,Moch:2017uml,Moult:2018jzp,Bruser:2018rad,Banerjee:2018ozf,Bizon:2018foh,Herzog:2018kwj,Henn:2019rmi,Bruser:2019auj,Henn:2019swt,Bacchetta:2019sam,Das:2019btv,vonManteuffel:2020vjv,Ebert:2020qef,Ebert:2020sfi,Agarwal:2021zft,Re:2021con,Lee:2021uqq,Duhr:2022yyp,Moult:2022xzt,Neumann:2022lft,Duhr:2022cob,Ju:2023dfa,Camarda:2023dqn,Moos:2023yfa,Baranowski:2024vxg,Billis:2024dqq,Benitez:2024nav,Aglietti:2025jdj,Benitez:2025vsp,Hoang:2025uaa,Jaarsma:2025tck,Buonocore:2026kgu} and even resummation of next-to-leading-power effects at leading-logarithmic accuracy~\cite{Moult:2018jjd,Moult:2019uhz,Beneke:2022obx}.
In parallel, increasingly sophisticated approaches have been developed to account for non-perturbative effects in several event shapes. 
These theoretical advances, together with improved analyses of LEP data and the prospects offered by future high-luminosity lepton colliders, place determinations of $\alpha_s$ from event shapes in a new regime of precision.

However, despite these favourable conditions, significant discrepancies persist among different determinations of $\alpha_s$ from event-shape observables~\cite{dEnterria:2022hzv}. 
These discrepancies concern not only the extracted central values of $\alpha_s$, but also the associated uncertainties and the interpretation of systematic effects. 
They largely reflect differing assumptions in the theoretical frameworks employed, including choices in perturbative matching and resummation, as well as in the treatment of hadronization and other non-perturbative power corrections. 

The current situation is suboptimal. 
Theory uncertainties in extractions of $\alpha_s$ based on analytic treatments of non-perturbative corrections, combined with higher-order resummation and fixed-order matching, are typically estimated at the level of around 1\%, reflecting the present sophistication and maturity of theoretical predictions. 
However, these extractions systematically yield lower central values than other determinations and than the current world average. 

The origin of these discrepancies remains poorly understood. 
More generally, there is still no broad consensus within the theory community on how to estimate uncertainties associated with a given theoretical framework.
On the one hand, this lack of agreement has led to a conservative decision: to exclude from the world average the determinations that employ analytic models of hadronization due to the lack of consensus on how to estimate their uncertainties and to the low extracted value of $\alpha_s$. 
This is illustrated in the latest PDG plot summarizing the $\alpha_s$ extractions that contribute to the world average shown on the right of \fig{PDG}. 
The excluded determinations nevertheless incorporate several state-of-the-art theoretical ingredients, both in perturbative resummation and in the field-theoretic treatment of hadronization effects. 
This situation raises concerns, on a more fundamental level, about the realization of a self-consistent precision QCD programme, with important implications both for the High-Luminosity LHC and, prospectively, for future electron-positron colliders.
In these environments, experimental uncertainties are expected to continue decreasing, and for many key observables the dominant limitation will increasingly come from theoretical systematics.

The lack of agreement among high-precision determinations based on different theoretical frameworks, each incorporating state-of-the-art ingredients, risks turning theoretical uncertainty into an effectively irreducible floor, not because the relevant effects are fundamentally unknowable, but because they are not yet being identified, standardized, and quantified within a common framework.
Rather than representing a retreat from the precision frontier, this situation should instead be viewed as a major opportunity. As uncertainties approach the percent level, effects that were previously negligible, such as long-distance contributions, matching choices, resummation prescriptions, and power corrections, become phenomenologically relevant and demand significantly closer scrutiny.
The objective is to disentangle which assumptions are responsible for shifts in central values, to systematically enumerate the missing ingredients in a given theoretical prediction, and to assign uncertainties that reflect clearly defined variations and parametric inputs. 

Addressing these issues calls for a coordinated programme that brings together theorists and experimentalists, establishes shared benchmarks and validation tests, and converges on a transparent and broadly accepted methodology for uncertainty estimation. It is worth recalling that lattice determinations themselves exhibited significant discrepancies in the past, which were progressively resolved through sustained community effort within the lattice community.
While lattice QCD determinations of $\alpha_s$ have now reached impressive precision~\cite{DallaBrida:2026kuo}, this does not diminish the importance of collider-based extractions. For a fundamental parameter of the Standard Model, robustness requires multiple independent and complementary methods. Moreover, $\alpha_s$ is a running coupling: lattice determinations are performed at energy scales far below those probed in high-energy collisions, and consistency across these scales provides a highly non-trivial test of QCD itself.

Precision physics is a cornerstone of the collider programme, where $\alpha_s$ enters ubiquitously, not only as a parameter to be measured, but also as an essential input to virtually all theoretical predictions. 
Future lepton colliders, in particular, will provide multiple clean and independent avenues for its extraction, thereby offering a unique environment to test the consistency of precision QCD at unprecedented accuracy.
We therefore see the establishment of a dedicated task force on $\alpha_s$ determinations from event shapes as a timely and necessary step for the success of the broader precision programme. 
The FCC Physics, Experiments and Detectors (PED) working group (WG) provides a natural framework for coordinating such an effort. The Task Force will also benefit from the support of the LHC Physics Center at CERN (LPCC). 
By fostering collaboration across experiments and theory groups, and by providing visibility, continuity, and institutional support, the FCC WG will play a central role in consolidating the community around a coherent long-term precision QCD strategy.

\section{Scope of the Task Force}
The Task Force on Strong Coupling Determinations from Event Shapes is conceived as a coordinated forum for assessing and improving determinations of $\alpha_s$ from event shape and collider correlators.
Its central role is to turn persistent differences among existing analyses into a concrete programme of benchmarks, validation tests, and targeted follow-up studies.
The goal is to foster a community effort to reach consensus on which assumptions underlying the various approaches are justified, to pinpoint where these assumptions require further validation, and to define the theoretical and experimental work needed to enable such tests in a way that can be applied consistently across observables and processes.

The task force is deliberately organized around specific technical questions for which progress requires sustained interaction between the groups involved.
This format is intended to complement standard conferences and workshops, where the relevant issues can be identified but usually cannot be resolved in sufficient detail.
The emphasis is on validation of assumptions, common inputs, controlled comparisons, repeated cross-checks, and documented follow-up work, so that disagreements between approaches can be traced to specific assumptions, different perturbative ingredients and missing higher orders, experimental inputs, or implementation choices.

\section{Kick-off Meeting: Strong coupling from Thrust at Lepton Colliders}
\begin{figure}[t]
\centering
\begin{minipage}[t]{0.48\textwidth}
\vspace{0pt}
\includegraphics[height=0.21\textheight,keepaspectratio]{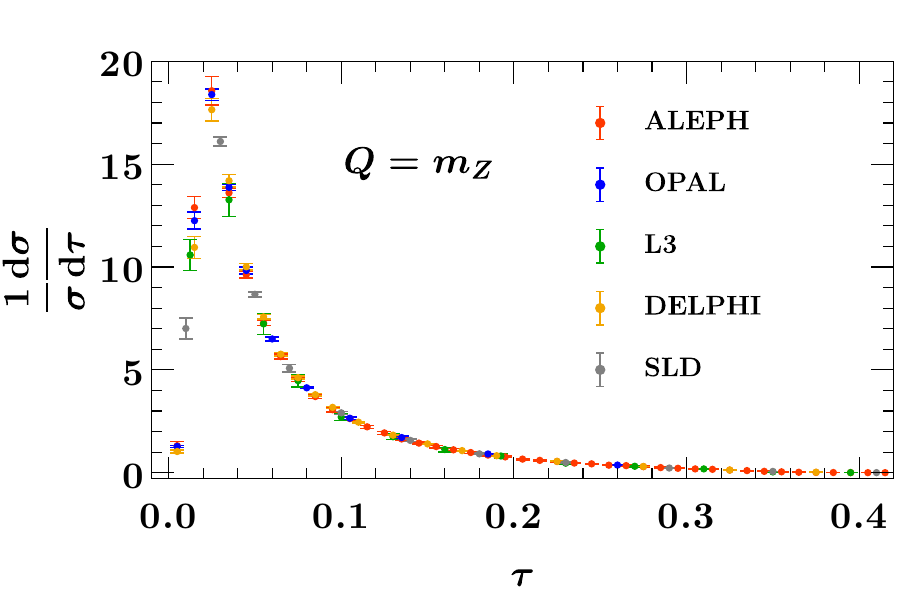}
\end{minipage}
\begin{minipage}[t]{0.48\textwidth}
\vspace{0pt}
\centering
\includegraphics[height=0.21\textheight,keepaspectratio]{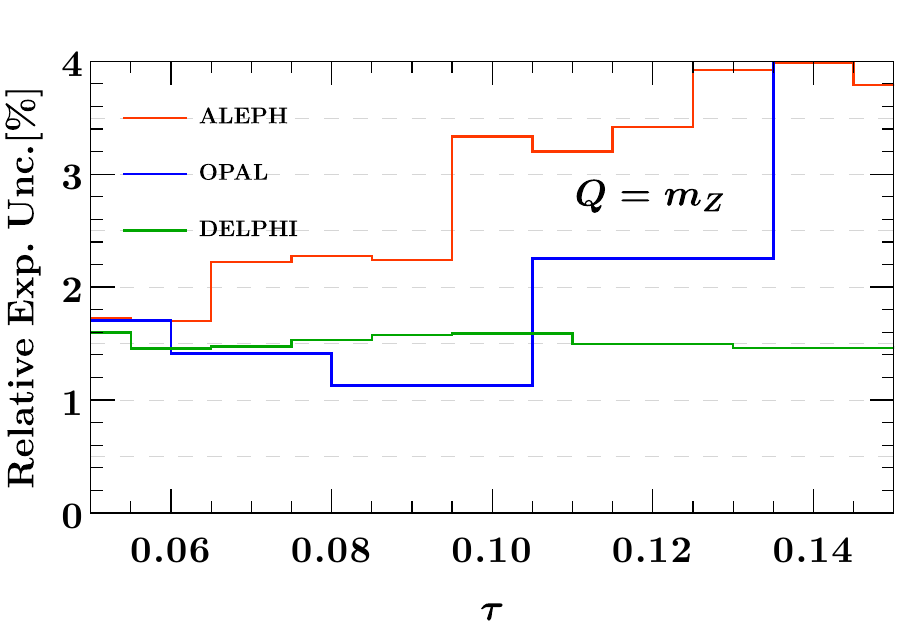}
\end{minipage}
\caption{Data at the Z pole for thrust are extremely precise. This level of precision, combined with data from multiple experiments (ALEPH, OPAL, L3, DELPHI, SLD)~\cite{ALEPH:2003obs,OPAL:2004wof,L3:2004cdh,DELPHI:2003yqh,SLD:1994idb}, provides a stringent test of theoretical predictions.
Left: Thrust distribution measured at the Z pole by LEP and SLD experiments~\cite{ALEPH:2003obs,OPAL:2004wof,L3:2004cdh,DELPHI:2003yqh,SLD:1994idb}. Right: Relative experimental uncertainties for thrust measurements from ALEPH, OPAL, and DELPHI~\cite{ALEPH:2003obs,OPAL:2004wof,DELPHI:2003yqh} in the $\tau \in [0.06,0.15]$ region. Uncertainties are at the 1--4\% level in this region.}
\label{fig:thrust_data}
\end{figure}

\label{sec:meeting}

The kick-off meeting of the LPCC Task Force took place on November 27--28, 2025, at CERN. The meeting was held in hybrid format, allowing for remote participation. It was attended by both theorists working on perturbative QCD, resummation, and non-perturbative effects, as well as experimentalists involved in LEP data analysis and associated FCC-ee projections, and members of the $e^+e^-$ alliance (EPA) who are currently working on LEP data reanalysis.

\subsection{Meeting structure}

A central motivation for establishing the task force was the recognition that the open questions in $\alpha_s$ extractions require a different format from standard conference discussions.
While talks on $\alpha_s$ determinations are regularly presented at both broad and specialized conferences (see, for example, \cite{Proceedings:2011zvx, Proceedings:2015eho, Proceedings:2019pra,alphas2022,alphas2025,SCET2024,PSR2025}), the typical post-talk Q\&A format does not provide sufficient time to thoroughly examine underlying assumptions, compare methodologies in detail, or identify the precise origins of discrepancies between different approaches.

For this reason, the workshop adopted a deliberately discussion-driven format. 
The theory sessions featured very brief presentations (about 10 minutes each, with minimal slides), followed by extended discussions lasting more than an hour per topic.
This structure proved highly effective: the discussions readily expanded to fill the allotted time, underscoring the need for in-depth exchanges of this kind, which are essential for progress on these technically subtle issues.

The workshop was organized around four main scientific themes:
\begin{itemize}
\item Day 1 (Morning): Impact of 3-jet power corrections.
\item Day 1 (Afternoon): Hadron mass effects and universality aspects of power corrections.
\item Day 1 (Afternoon): Experimental measurements and new analyses of LEP data.
\item Day 2: Impact of resummation space choice and perturbative uncertainties.
\end{itemize}

The experimental session followed a different but equally effective format: three presentations spread over two hours, with extensive interaction and frequent interruptions from theorists seeking to clarify specific aspects of the analyses. The questions and ensuing discussions naturally filled the entire session, highlighting the value of allowing ample time for detailed exchanges between theory and experiment.

The full agenda and all presentation materials are available on the Indico page~\cite{indico}.
In what follows, each topic is summarized by distinguishing between the key points presented at the meeting and the conclusions that emerged from the ensuing discussions.

\subsection{Experimental measurements and new analyses of LEP data}

\subsubsection{LEP data reanalyses}

The EPA aims to curate and analyze archived $e^+e^-$ collider data to enable new precision studies. At this workshop, preliminary thrust reanalyses of archived ALEPH and DELPHI data were presented. 
The goal of these measurements is to measure the thrust spectrum with fine binning, extending deep into the peak region where non-perturbative effects are significant and were not previously explored in detail, and to understand the experimental effects that drive the central values and uncertainties of $\alpha_s$ extractions from thrust.

\begin{figure}[ht]
\begin{center}
\includegraphics[width=0.7\textwidth]{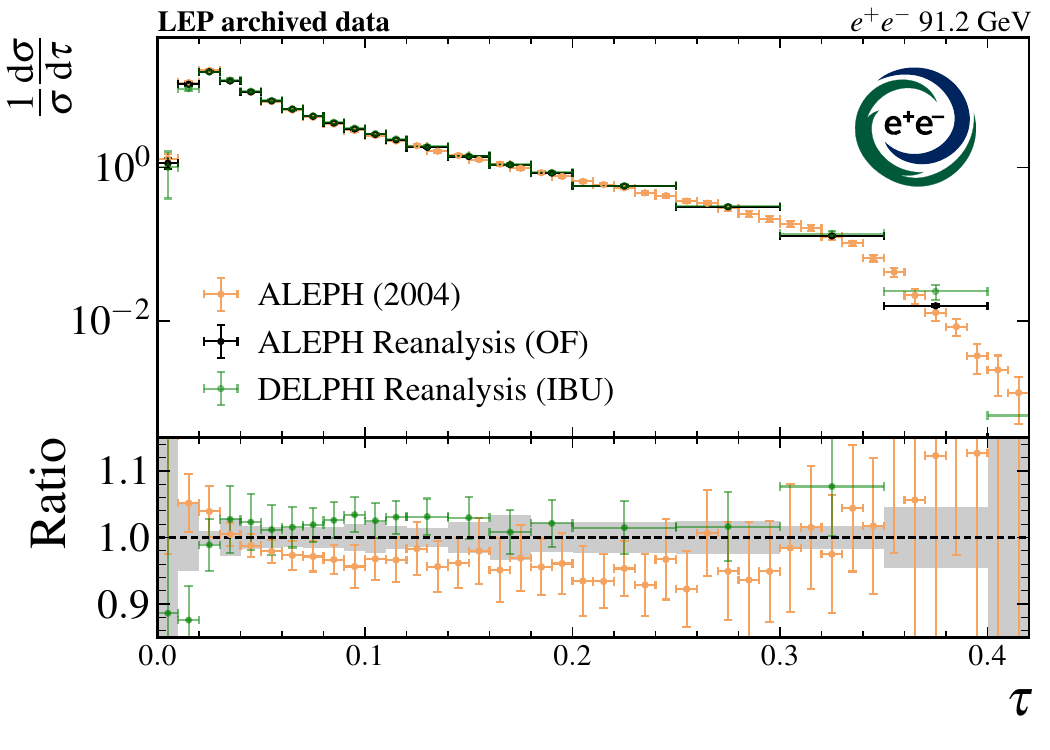}
\caption{Comparison of previous ALEPH 2004 publication (orange) with the ALEPH (black) and DELPHI (green) preliminary reanalyses. The ALEPH reanalysis utilizes the unbinned OmniFold (OF) unfolding method~\cite{Andreassen:2019cjw}, while the DELPHI reanalysis utilizes the binned Iterative Bayesian Unfolding (IBU) method~\cite{DAgostini:2010hil}. The top panel shows the differential cross section spectrum, while the bottom panel shows the ratio to the ALEPH reanalysis. The indicated error bars are the total uncertainty from statistical, experimental, and theoretical components. A systematic shift is observed with respect to the previous ALEPH measurement. While many checks were already performed, further studies are ongoing to understand the origin of the shifts.}
\label{fig:lep_reanalysis}
\end{center}
\end{figure}

The preliminary reanalysis of ALEPH data~\cite{Electron-PositronAlliance:2025hze} shows a systematic shift in the thrust distribution compared to the previous ALEPH measurement~\cite{ALEPH:2003obs} as shown in \fig{lep_reanalysis}. 
The preliminary reanalysis of DELPHI data~\cite{Zhang:2025nlf}, presented for the first time in comparison with theoretical predictions at this workshop, shows an even stronger systematic shift that is consistent in direction with the ALEPH reanalysis findings. Note that the ALEPH reanalysis is unbinned through the use of the OmniFold (OF) unfolding method~\cite{Andreassen:2019cjw}, so the binning is chosen at the plotting stage based on which reference is compared to. The DELPHI reanalysis is binned and uses the D'Agostini Iterative Bayesian Unfolding (IBU) method~\cite{DAgostini:2010hil}. Since the results presented were preliminary and a major goal was to collect feedback on the analysis procedure from other LEP experts, dedicated fits to the spectrum were not yet carried out and will be performed once the final results are made public.

\begin{figure}[ht]
\begin{center}
\includegraphics[width=0.48\textwidth]{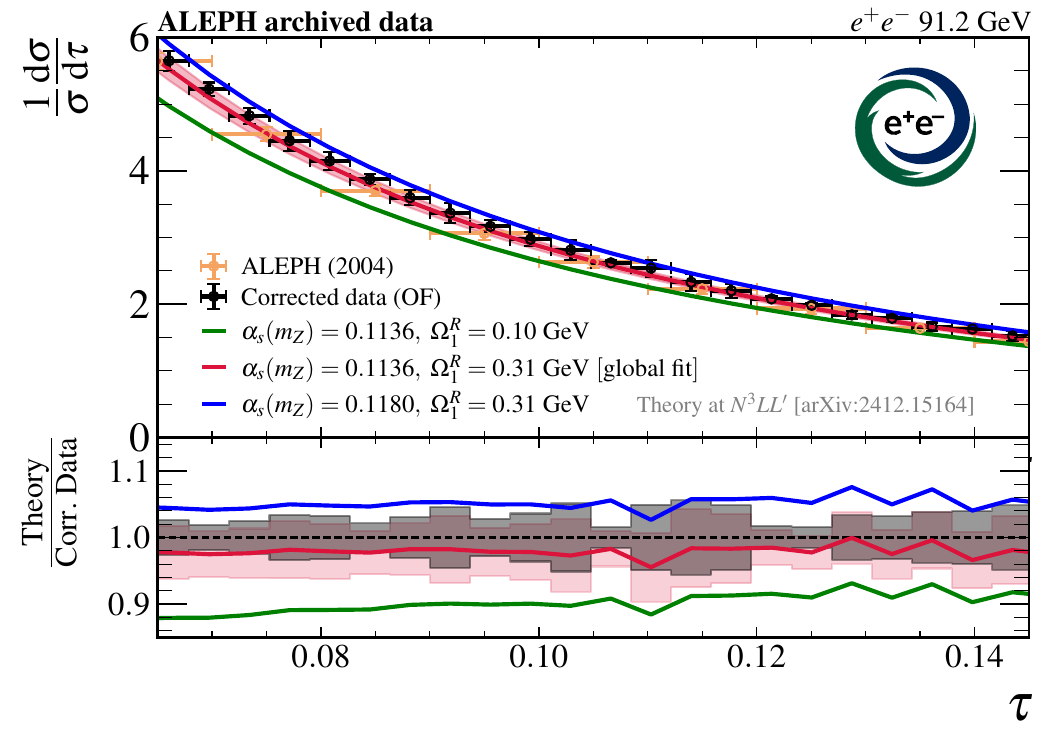}
\includegraphics[width=0.48\textwidth]{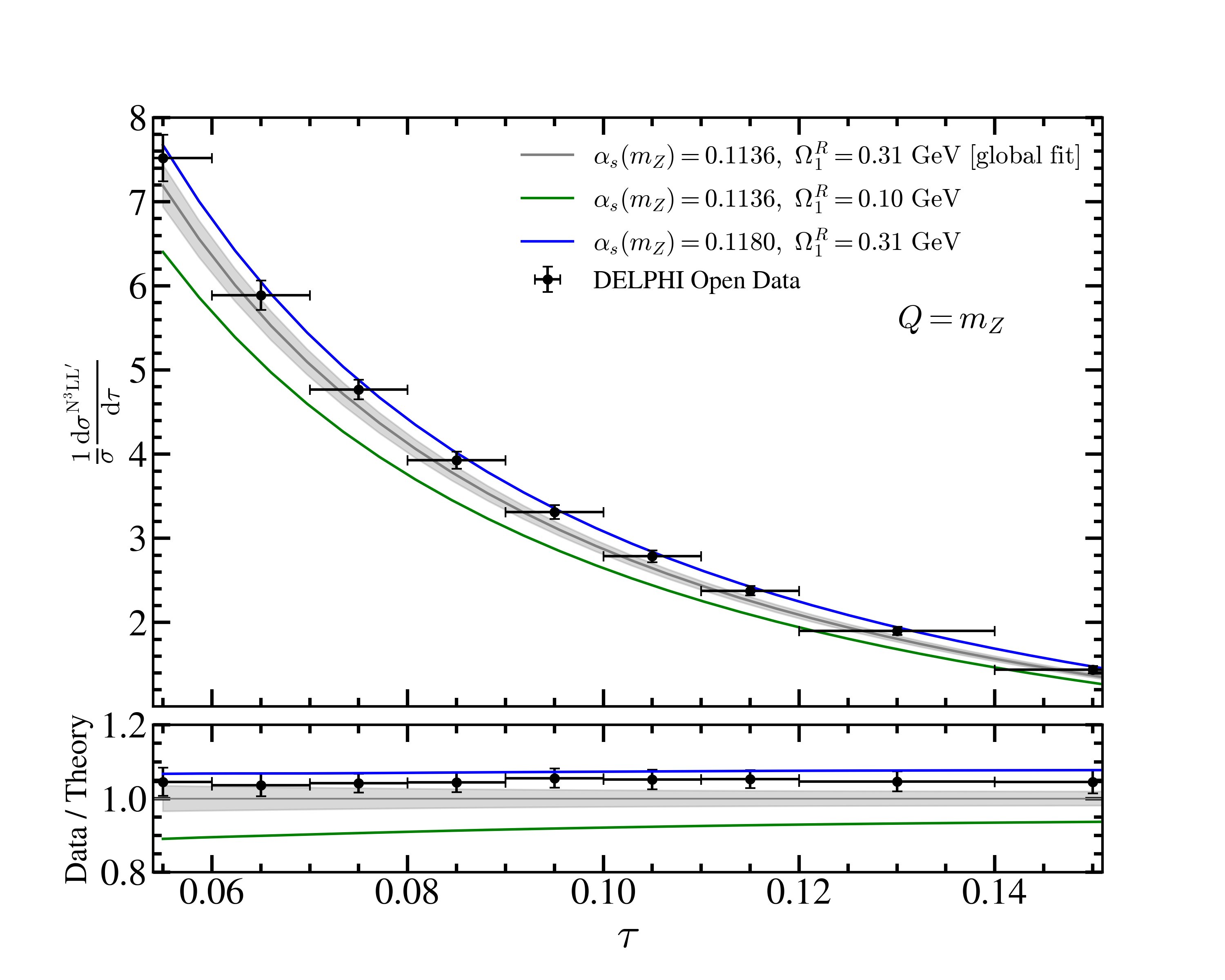}
\caption{Comparison of the preliminary ALEPH (left) and DELPHI (right) reanalyses to theoretical predictions at N$^3$LL$^\prime$ accuracy. 
The observed shifts are shown to be comparable in size to the difference between N$^3$LL$^\prime$ predictions from Ref.~\cite{Benitez:2024nav} using $\alpha_s(m_Z)=0.1180$ and $\alpha_s(m_Z)=0.1136$ at fixed $\Omega_1^R = 0.31$~GeV. Note that similar shifts can also be obtained with a variation of $\Omega_1^R$, illustrating the well-known anti-correlation between $\alpha_s$ and $\Omega_1$ in single observable distributions at a fixed collision energy. 
These results motivate a detailed understanding of the differences with respect to previous LEP results and potentially new fits with the reanalyzed data. }
\label{fig:lep_reanalysis_theoryComp}
\end{center}
\end{figure}

Even without a new fit, it can be seen in \fig{lep_reanalysis_theoryComp} that the shifts shown are comparable in size to the difference between N$^3$LL$^\prime$ predictions from Ref.~\cite{Benitez:2024nav} using $\alpha_s(m_Z)=0.1180$ and $\alpha_s(m_Z)=0.1136$ at fixed $\Omega_1^R = 0.31$~GeV, the parameter controlling the leading non-perturbative effect. On the other hand, the same figure shows that a similar shift can be obtained with a variation of $\Omega_1^R$, demonstrating the well-known degeneracy problem between $\alpha_s$ and $\Omega_1$ in the fit of the strong coupling with a single observable and a single collision energy.
These studies indicate that effects on $\alpha_s$ fits can be quantitatively sizable, motivating a detailed understanding of the differences with respect to previous LEP results and eventually new fits using the reanalyzed data.

\subsubsection{Discussion outcomes}

The session resulted in a constructive exchange between theory and experiment regarding systematic uncertainties and the practical requirements for enabling precise, reproducible comparisons between data and state-of-the-art theory predictions.

It was clear from the discussions that the availability of reanalyzed datasets with modern unfolding techniques and properly documented correlations would be invaluable for addressing long-standing discrepancies between $\alpha_s$ extractions and for providing robust test cases to benchmark theoretical developments. 
Key points of agreement include the need for a priority list of observables beyond thrust, the importance of correlation matrices across observables, and the value of extending reanalyses to multiple center-of-mass energies to test the $Q$-dependence of power corrections.

It was also agreed that the priority is to understand the origin of the differences between the reanalyses and the original ALEPH and DELPHI publications, including the dissenting DELPHI results in the Daniel Wicke thesis work~\cite{Wicke:1999zz}.
While many checks were already performed, additional studies were discussed. Beyond the use of modern unfolding methods, several experimental procedures differ from those used in the original analyses. 
Some of these differences reflect updated understanding of theoretically well-defined observables. 
For example, in the previous ALEPH result~\cite{ALEPH:2003obs} the mismodeling of neutral particles in Monte Carlo (MC) generation and detector response was accounted for by removing all neutrals in a selected energy range, recomputing thrust, and taking the difference with respect to the nominal as a systematic. Such a systematic effectively redefines the observable from all-particle thrust to charged-particle thrust, which has itself come to be recognized as an interesting observable since the LEP era~\cite{Chang:2013iba}. For the reanalysis, an alternative approach was therefore adopted to vary the neutral-particle efficiency and scale rather than excluding neutrals altogether. This procedure and other differences from the previous LEP analyses are outlined in the EEA papers, notes, and slides~\cite{EEAlliancePapers}.

Other differences arise from current limitations of the archived data. 
Some examples of such limitations include the absence of dedicated MC samples without initial-state radiation, which were used by ALEPH for these corrections, or the restriction to the 1994 data-taking period due to the availability of reconstructed MC only for the corresponding detector conditions. 
These limitations will be addressed in collaboration with the LEP experimental community through additional data curation, and their impact will be carefully evaluated and documented.

\subsection{2-jet versus 3-jet power corrections}
A concern raised in recent discussions of strong-coupling determinations based on analytic models of power corrections is the possible impact of non-perturbative effects beyond the dijet limit~\cite{Luisoni:2020efy,Caola:2021kzt,Caola:2022vea}; this concern was one of the primary reasons for excluding these determinations from the 2023 PDG average~\cite{ParticleDataGroup:2024cfk}.

Modern matched predictions do include the contribution associated with 3-jet kinematics at the perturbative level; they retain the full fixed-order result, including non-singular terms, whose numerical impact in commonly used fit regions is not negligible. On top of this perturbative description, non-perturbative effects beyond the leading-power dijet picture may induce an additional $\tau$ dependence that affects the extraction of $\alpha_s$.

Non-perturbative corrections to the thrust distribution are, \emph{a priori}, a complicated function of $\tau$.
In the dijet regime ($\tau \ll 1$), however, a significant simplification occurs.
At leading power in the $\tau \ll 1$ expansion, one can show~\cite{Lee:2006nr,Fleming:2007qr,Schwartz:2007ib,Abbate:2010xh} that non-perturbative effects with scaling such as $\cO((\lqcd/(Q\tau))^n)$ can be organized within a factorization theorem involving a universal soft function.
Among these, the leading non-perturbative correction in the region $\lqcd \ll Q \tau \ll Q$ is given by the linear term, $\cO(\lqcd/(Q\tau))$, which is parametrized by a $\tau$-independent non-perturbative parameter~\cite{Korchemsky:1994is,Korchemsky:1999kt,Lee:2006nr,Hoang:2007vb,Abbate:2010xh}
\begin{equation}
\label{eq:MSbarOmega1}
\Omega_1 = \frac{1}{N_c} \langle 0 | \text{tr} ~ \overline{Y}_{\bar{n}}^T(0) Y_n (0) \hat {\cal E}_T(0) Y_n^{\dagger}(0)\overline{Y}_{\bar{n}}^{*}(0) | 0 \rangle\,.
\end{equation}
Analogous results can be obtained from renormalon analysis and effective coupling models~\cite{Dokshitzer:1995qm,Dokshitzer:1997iz,Dokshitzer:1998pt,Gardi:2001ny,Gardi:2003iv,Berger:2004xf}, in which case the non-perturbative parameter parametrizing the $\cO(\lqcd/(Q\tau))$ corrections reads
\beq
\alpha_0(\mu_I) = \frac{1}{\mu_I} \int_0^{\mu_I} \df \mu \, \alpha_s(\mu)\,,
\eeq
where $\mu_I$ is an infrared matching scale separating perturbative and non-perturbative dynamics.
For the present discussion, the distinction between the effective coupling and the field-theoretic matrix element is not crucial, and we will use $\alpha_0$ and $\Omega_1$ interchangeably.

This dijet picture breaks down at larger values of thrust, where $\tau$-suppressed non-perturbative effects probing the geometry of the multi-jet configuration can become important.
The power correction then acquires a more complex, $\tau$-dependent structure, commonly parametrised by a function $\zeta(\tau)$, with $\zeta(\tau) \to \text{const}$ in the dijet limit.
While current techniques remain far from enabling a first-principles calculation of $\zeta(\tau)$ in full QCD, phenomenological models across the thrust spectrum have been explored in several works (see, e.g., \refcite{Caola:2022vea}), including detailed investigations of their impact on event-shape $\alpha_s$ fits, as pioneered in \refscite{Nason:2023asn,Nason:2025qbx}.

\begin{figure}[ht]
\begin{center}
\begin{minipage}[t]{0.45\linewidth}
\vspace{0pt}
\centering
\includegraphics[height=0.22\textheight,keepaspectratio]{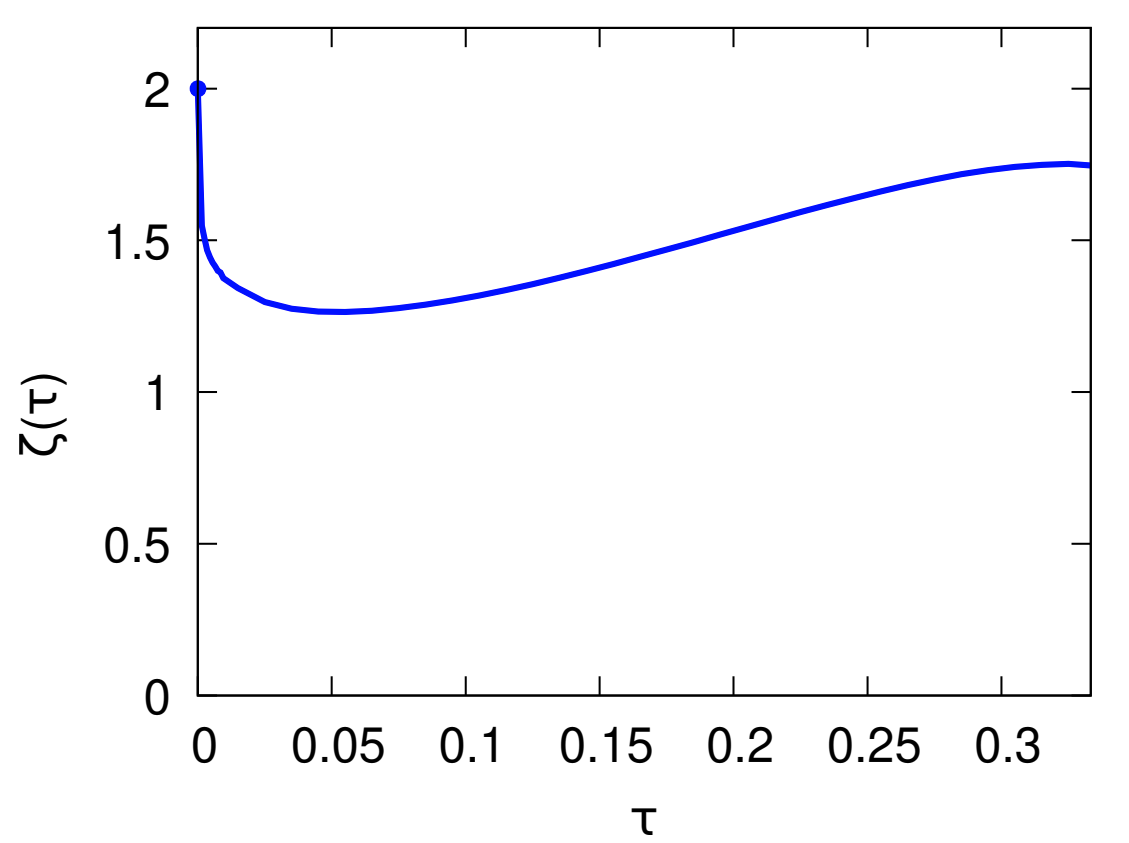}
\end{minipage}\hspace{-0cm}
\begin{minipage}[t]{0.45\linewidth}
\vspace{0pt}
\centering
\includegraphics[height=0.22\textheight,keepaspectratio]{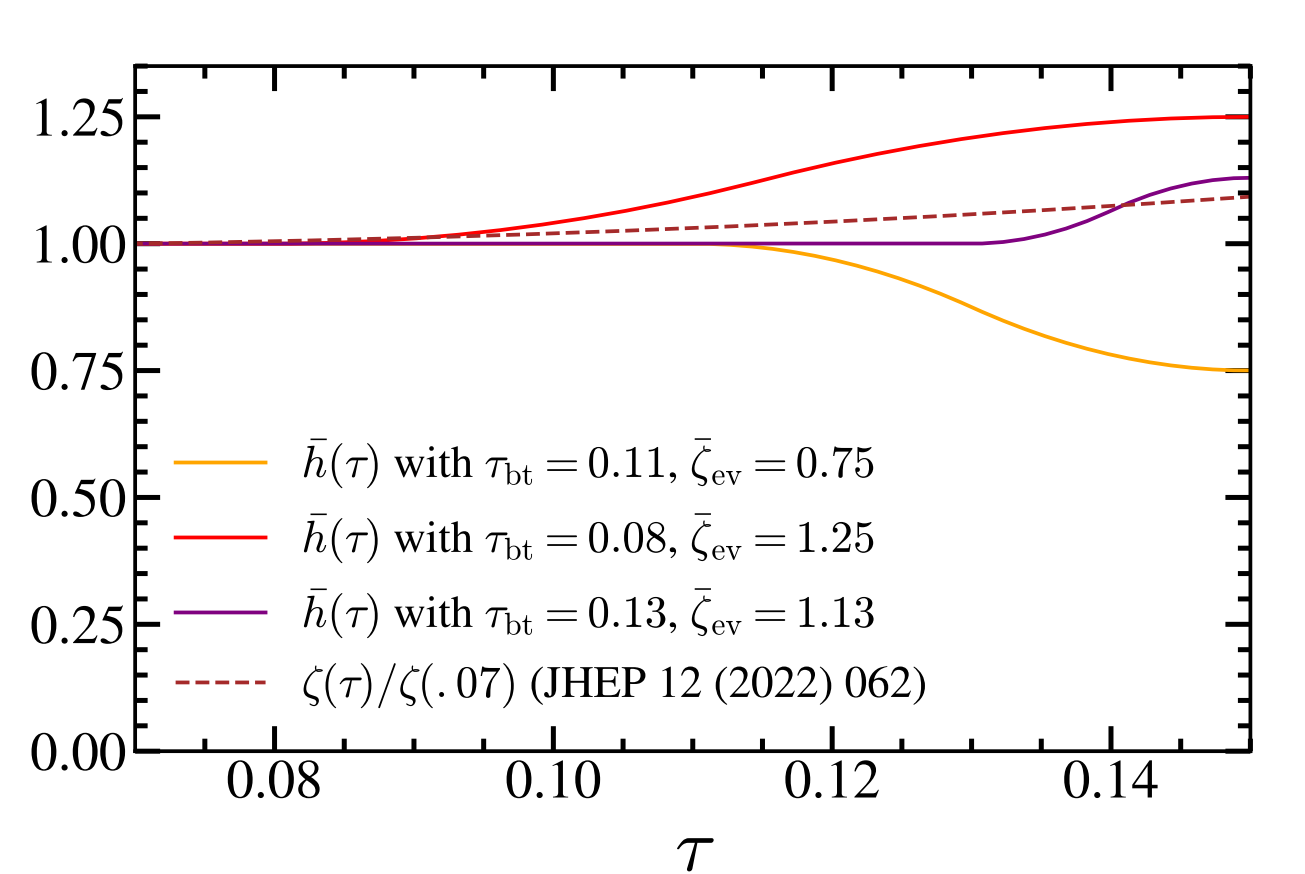}
\end{minipage}
\caption{{\bf Left:} Plot from \refcite{Nason:2023asn} of the 3-jet power correction model $\zeta(\nu)$ of \refcite{Caola:2022vea}. {\bf Right:} Some 3-jet power corrections models considered in \refcite{Benitez:2024nav} in the fit region $\tau \in [0.07, 0.15]$. The two panels have different ranges on the $x$-axis and different normalizations on the $y$-axis. To compare them, note that the blue line in the left panel corresponds to the dashed red line in the right panel.}
\label{fig:ztau}
\end{center}
\end{figure}

The model employed in these analyses estimates the full set of non-perturbative power corrections at a given $\tau$ using an Abelian approximation, in which a hard final-state gluon is replaced by a photon and a soft gluon with a small mass $\lambda$ is emitted from the resulting $q\bar{q}\gamma$ three-jet–like configuration.
For thrust, this model yields a functional form, $\zeta(\tau)$, which parametrizes the three-jet power correction. This function is shown in the left panel of \fig{ztau}. The plot shows that the size of power corrections varies significantly when going from the dijet limit to the region used in the fits. 

A complementary phenomenological approach is proposed in \refcite{Benitez:2024nav}, which introduces a function $\bar h(\tau)$ to encode the deviation of the non-perturbative correction from a constant value. This function is assumed to be nearly constant at the beginning of the fit region ($\tau=0.05$) and has parameters that can be varied to reproduce a range of possible deviations towards the end of the fit region ($\tau=0.15$), including variations that encompass the model of \refcite{Caola:2022vea} (dashed brown curve), as shown in the right panel of \fig{ztau}.

Before concluding this section, we want to stress that both models based on $\Omega_1$ and $\alpha_0$ only account for power corrections that are linear in $\lqcd$ and therefore are valid in a regime where $Q \tau \gg \lqcd$. 
As $Q \tau$ approaches the hadronization scale ($\mu_I\sim 1-2\,\mathrm{GeV}$), power corrections with scaling $\cO(\lqcd^n/(Q\tau)^n)$ become of comparable size to the linear ones, and, despite their increased powers in $\lqcd$, cannot be neglected.
In both formalisms these corrections can be accounted for by including the full non-perturbative information about the shape function or the effective coupling, or equivalently by including (multiple) higher moments of either of them.
In either case, this comes with a large increase in the number of non-perturbative parameters and it is one of the primary reasons why many event-shape fits explicitly exclude the peak region despite the abundance of data in that range.
It should then be clear that the 3-jet power corrections discussed in this section by no means account for this point, but rather focus on the $\cO(\lqcd/Q)$ non-perturbative effects stemming from the 3-jet region.

\subsubsection{Presentation summary}
\begin{figure}[ht]
\centering
\begin{minipage}[t]{0.43\textwidth}
\vspace{0pt}
\centering
\includegraphics[width=\linewidth]{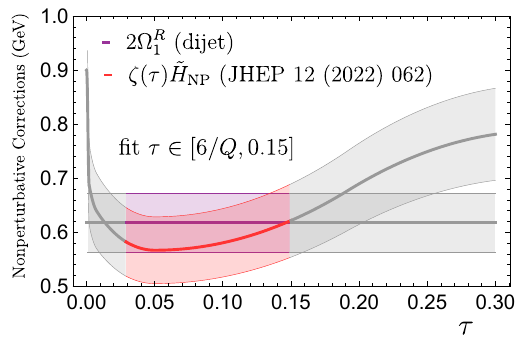}
\end{minipage}
\hfill
\begin{minipage}[t]{0.55\textwidth}
\vspace{0pt}
\centering
\scriptsize
\setlength{\tabcolsep}{2.8pt}
\renewcommand{\arraystretch}{0.95}
\resizebox{\linewidth}{!}{%
\begin{tabular}{|l||c|c||c|c||c|c||c|c||}
    \hline
     & \multicolumn{8}{|c|}{$\as(M_Z)$} \\
    \hline
     & \multicolumn{2}{|c||}{CTy3} &  \multicolumn{2}{|c||}{C} &  \multicolumn{2}{|c||}{T} &  \multicolumn{2}{|c||}{$y_3$} \\
    \hline
 Variation & $\zeta(v)$ & $\zeta(0)$ & $\zeta(v)$ & $\zeta(0)$ & $\zeta(v)$ & $\zeta(0)$ & $\zeta(v)$ & $\zeta(0)$ \\
\hline
default & 0.1181 & 0.1161 & 0.1169 & 0.1139 & 0.1168 & 0.1158 & 0.1155 & 0.1154 \\
\hline
$\mu_R=\mu_0/2$ & 0.1167 & 0.1155 & 0.1141 & 0.1105 & 0.1159 & 0.1128 & 0.1122 & 0.1131 \\
\hline
$\mu_R=2\mu_0$ & 0.1167 & 0.1150 & 0.1212 & 0.1184 & 0.1208 & 0.1191 & 0.1157 & 0.1161 \\
\hline
std scheme & 0.1173 & 0.1153 & 0.1164 & 0.1118 & 0.1152 & 0.1148 & 0.1150 & 0.1149 \\
\hline
p scheme & 0.1160 & 0.1141 & 0.1164 & 0.1118 & 0.1152 & 0.1148 & 0.1137 & 0.1135 \\
\hline
D scheme & 0.1199 & 0.1173 & 0.1190 & 0.1153 & 0.1205 & 0.1170 & 0.1168 & 0.1166 \\
\hline
$C_{\rm ll}=1.5$ & 0.1165 & 0.1143 & 0.1151 & 0.1116 & 0.1154 & 0.1133 & 0.1142 & 0.1142 \\
\hline
$C_{\rm ll}=3$ & 0.1177 & 0.1159 & 0.1221 & 0.1116 & 0.1180 & 0.1172 & 0.1156 & 0.1154 \\
\hline
non-pert scheme (b) & 0.1193 & 0.1163 & 0.1191 & 0.1176 & 0.1185 & 0.1184 & 0.1154 & 0.1154 \\
\hline
non-pert scheme (c) & 0.1189 & 0.1167 & 0.1195 & 0.1172 & 0.1192 & 0.1191 & 0.1154 & 0.1154 \\
\hline
minus non-pert error & 0.1187 & 0.1161 & 0.1173 & 0.1139 & 0.1165 & 0.1158 & 0.1157 & 0.1154 \\
\hline
plus non-pert error & 0.1189 & 0.1161 & 0.1172 & 0.1139 & 0.1172 & 0.1158 & 0.1153 & 0.1154 \\
\hline
\end{tabular}}
\end{minipage}
\caption{{\bf Left:} Fit results for $2\Omega_1^R$ and for $\zeta(\tau)\,\tilde H_{\rm NP}$ as a function of $\tau$ using data in the range $\tau \in [(6\,{\rm GeV})/Q,0.15]$, from \refcite{Benitez:2024nav}. The colored band indicates the fit window, within which the behavior of the power correction is consistent for the different treatments. {\bf Right:} Table~7 of \refcite{Nason:2025qbx}, summarizing the fitted values of $\alpha_s(M_Z)$ obtained with the event-shape-dependent model $\zeta(v)$ and with the constant-shift limit $\zeta(0)$ for $C$, $T$, $y_3$ and the joint CTy3 fits under the variations considered in that work.}
\label{fig:power_corrections}
\end{figure}

At the workshop, G. Zanderighi presented the results of \cite{Nason:2025qbx}.
The study uses fixed-order perturbation theory to $\cO(\alpha_s^3)$ in the three-jet region, without resummation, and performs $\alpha_s$ fits in several different setups.
A central element of the presentation was the comparison between two assumptions for the power correction: a dijet model $\zeta(0)$, corresponding to a constant shift, and an event-shape-dependent 3-jet model $\zeta(\nu)$ based on \refcite{Caola:2022vea}, applied to several event-shape observables.
In all fits performed, joint thrust, $C$-parameter, and $y_{3}$ fits, or fits to individual variables, the 3-jet model was found to give systematically larger values of $\alpha_s$, with differences ranging from a few per-mille to several percent as shown in the table reported in the right panel of \fig{power_corrections}.

V. Mateu, a coauthor of \refcite{Benitez:2024nav}, presented a complementary perspective based on the implementation in that work of the three-jet model of \refcite{Caola:2022vea}. Since it was observed that the three-jet correction varies only mildly across the fit region used in their analysis, it is argued that the primary effect is an overall normalization of the non-perturbative contribution. 
Therefore, given that this contribution is itself fitted to data, such a normalization should be largely absorbed into the extracted value of $\Omega_1$, with a correspondingly smaller impact on the determination of $\alpha_s$. 
In particular, it was highlighted that the $\sim$10--15\% uncertainty on the dijet power-correction parameter $\Omega_1$ in the two-dimensional $(\Omega_1,\alpha_s)$ fit is already large enough that a three-jet correction with such mild $\tau$ dependence is effectively indistinguishable, within uncertainties, from a constant dijet correction, as shown in the left panel of \fig{power_corrections}.

This point was further illustrated in \refcite{Benitez:2024nav} by performing thrust fits with alternative functional forms for the 3-jet correction, spanning variations of up to about 25\%, and producing only very small effects on the fitted value of $\alpha_s$. 
This was attributed to the limited fit window and the intrinsic uncertainty on $\Omega_1$: the $\tau$ dependence of the 3-jet correction enters only as a small modification of an already subleading non-perturbative contribution.
A separate point emphasized in the presentation was that the validity of the dijet picture should first be assessed perturbatively, through a comparison between singular and non-singular contributions.
	
\subsubsection{Discussion outcomes}

No consensus was reached at the workshop on a sharp separation between dijet and 3-jet effects in the thrust fit region.
In fact, much of the discussion moved away from the specific size of 3-jet power corrections in thrust and toward the more basic issue of how the dijet and 3-jet regimes should be distinguished already at the perturbative level.
Several interventions questioned whether a genuinely dijet-dominated region extends into the phenomenologically relevant fit window at all, or whether it may be restricted to substantially smaller values of $\tau$, well below the region usually employed in thrust fits.
This question bears directly on the role of resummation, on the interpretation of singular and non-singular contributions, and on the appropriate starting point for modelling non-perturbative effects.

On the non-perturbative sector, the discussion indicated that such effects beyond the dijet limit can have a sizable impact in fits for event-shape observables, especially when extending the fit to include the region of larger values of the event-shape variable.

For thrust, assuming the model of~\refcite{Caola:2022vea} and the restricted fit window of~\refcite{Benitez:2024nav}, it was observed that the dominant effect of the three-jet correction can, to a large extent, be absorbed into the fitted value of $\Omega_1$, with a correspondingly smaller impact on the extracted value of $\alpha_s$ (see also columns~5--6 in the right panel of \fig{power_corrections}). However, it remains unclear whether the range of variations considered in the right panel of \fig{ztau} is sufficient to provide a reliable estimate of the associated uncertainty.

\noindent The discussion identified the following concrete follow-up studies:
\begin{itemize}
\item As a way to clarify the perturbative separation between the dijet and three-jet regimes, study the non-singular partonic cross section relative to the total partonic cross section order by order. In particular, assess whether the observed size of about 10\% at $\tau=0.1$ and 16\% at $\tau=0.15$ is stable across perturbative orders or instead an accidental feature of the $\cO(\alpha_s^3)$ result. 
Similar exercises should be repeated for other observables in which resummed predictions play an important role, such as other event shapes, as well as LHC observables like the transverse momentum of a color singlet and cross sections with vetoes on the leading-jet transverse momentum.
\item Compare thrust fits in the transition region---for instance N$^3$LL$^\prime$+$\cO(\alpha_s^3)$ fits in $\tau \in [0.15,0.33]$---with the dijet fits in $\tau \in [0.06,0.15]$, in order to assess whether the treatment of the transition from resummation to fixed order yields compatible results.
\end{itemize}

More broadly, the separation between dijet and 3-jet regimes, and the related question of when resummation is appropriate, necessary, or potentially detrimental to the accuracy of a prediction, emerged as an issue that extends well beyond the specific problem of $\alpha_s$ extractions from thrust and clearly requires further discussion.
This is a recurrent theme across collider phenomenology, with analogues in observables such as the $Z$-boson transverse momentum spectrum, jet-radius logarithms, and other multi-scale problems.
Therefore, developing clearer diagnostics and quantitative tests to assess different perturbative descriptions, their domains of validity, and the corresponding uncertainty estimates, in a way that can be applied consistently and universally across observables and processes is an important goal of the task force in the long term.

\subsection{Hadron mass effects}
Another source of non-perturbative uncertainty arises from the fact that event-shape observables are measured using hadron momenta, whereas perturbative calculations are typically formulated at the level of massless partons. 
For massless final-state particles, the energy and the magnitude of the three-momentum are identical, so different definitions of an observable that employ one or the other lead to the same result. 
For massive hadrons, however, these definitions no longer coincide and can induce sizable power corrections, particularly for soft particles~\cite{Salam:2001bd}. Indeed, while mass effects from energetic, collinear hadrons are suppressed as $\cO(\lqcd^2/(Q^2\tau))$, the contributions from soft hadrons with momenta of order $\lqcd$ enter at $\cO(\lqcd/(Q\tau))$~\cite{Salam:2001bd,Mateu:2012nk}, the same order as the leading non-perturbative correction itself, and can furthermore be logarithmically enhanced by multiple soft emissions~\cite{Salam:2001bd}. 

The treatment of hadron masses in the observable definition therefore directly affects the extraction of $\Omega_1$ and, consequently, the determination of $\alpha_s$. Modifications of the observable definition that leave it unchanged in the massless limit, but lead to distinct results in the presence of finite hadron masses, define different mass schemes.
A possible way to assess the impact of hadron-mass effects is therefore to compare results obtained using different schemes. 
Denoting the four-momenta of final-state particles by $p_i$, we list the mass schemes introduced and studied in \refcite{Salam:2001bd}:
\begin{itemize}
\item \textbf{E-scheme:} the spatial components of the momenta are rescaled as $\vec{p}_i \to p_i^0 \vec{p}_i / |\vec{p}_i|$;
\item \textbf{p-scheme:} the time component of the momenta is replaced by $p_i^0 = |\vec{p}_i|$ before computing the observable;
\item \textbf{D-scheme:} all particles are decayed isotropically into pairs of massless pseudoparticles, and the shape variables are evaluated using these pseudoparticles.
\end{itemize}

The impact of these scheme choices on the extraction of the non-perturbative parameters, such as $\Omega_1$, or $\alpha_0$, and $\alpha_s$ is, however, not universal. 
It depends on the analysis strategy, in particular on how the experimental measurement is defined or corrected, on the theoretical assumptions used in the theory modeling as well as on other aspects of the analysis such as the fit range. 
For this reason, differences between schemes should be interpreted in the context of the specific fit setup employed. These issues will be discussed in more detail in the next section.

\subsubsection{Presentations at the Workshop}

The issue of hadron masses was discussed at the workshop by P.~Nason and I.~Stewart, who presented two complementary perspectives on the topic.

P.~Nason reported on the study of hadron-mass effects in the fits of~\refscite{Nason:2023asn,Nason:2025qbx}. The results include a comprehensive set of fits to individual event-shape observables, in particular thrust, the C-parameter, and $y_{3}$, as well as simultaneous fits to combinations of two or three observables.
Their approach assesses the impact of different mass schemes using MC generators, with the only assumption being that the generator provides a faithful representation of the data. 
Since the perturbative calculation is performed for massless partons, it is independent of the mass scheme, allowing the scheme effects to be implemented entirely at the level of the data.

The strategy consists of generating a large sample of $e^+e^- \to \text{hadrons}$ events with \textsc{Pythia\,8}~\cite{Sjostrand:2014zea}. 
For each event, the observable is computed both in the standard experimental scheme and in an alternative scheme, $S = E$, $p$, or $D$.
This produces a migration matrix $T^{(S)}_{ij}$, where the entry $(i,j)$ counts the number of events that fall into bin $i$ in the standard scheme and into bin $j$ in the alternative scheme $S$. 
The experimental data are then converted to the alternative scheme according to
\begin{equation}
n_j^{(S)} = \sum_i n_i^{(\rm Std)} \frac{T^{(S)}_{ij}}{\sum_k T^{(S)}_{ik}}\,.
\end{equation}

A general conclusion from both single- and multiple-observable fits across all LEP energies is that the choice of hadron-mass scheme was found to be the dominant source of uncertainty. For example, the combined fit of thrust, C-parameter and $y_{3}$ yields $\alpha_s(m_Z) = 0.1181$ in the E-scheme (default), $\alpha_s(m_Z) = 0.1160$ in the p-scheme, and $\alpha_s(m_Z) = 0.1199$ in the D-scheme (see Table~\ref{tab:allfits2j}). 
The resulting spread of approximately $\pm 0.002$ is significantly larger than the quoted fit uncertainties in~\refscite{Abbate:2010xh,Benitez:2024nav}, raising the question of whether it reflects an unaccounted systematic.
\begin{table}[htb]
  \begin{center}
    {\small
\begin{tabular}
{|l||c|c||c|c||c|c||c|c||}
    \hline
     & \multicolumn{8}{|c|}{$\as(M_Z)$} \\
    \hline
     & \multicolumn{2}{|c||}{CTy3} &  \multicolumn{2}{c||}{C} &  \multicolumn{2}{c||}{\bf{T}} &  \multicolumn{2}{c||}{$y_3$} \\
    \hline
 Variation & $\zeta(v)$ & $\zeta(0)$ & $\zeta(v)$ & $\zeta(0)$ & $\boldsymbol{\zeta(v)}$ & $\boldsymbol{\zeta(0)}$ & $\zeta(v)$ & $\zeta(0)$ \\
\hline
$E$ scheme & 0.1181 & 0.1161 & 0.1169 & 0.1139 & \textbf{0.1168} & \textbf{0.1158} & 0.1155 & 0.1154 \\
\hline
No Variation & 0.1173 & 0.1153 & 0.1164 & 0.1118 & \textbf{0.1152} & \textbf{0.1148} & 0.1150 & 0.1149 \\
\hline
$p$ scheme & 0.1160 & 0.1141 & 0.1164 & 0.1118 & \textbf{0.1152} & \textbf{0.1148} & 0.1137 & 0.1135 \\
\hline
D scheme & 0.1199 & 0.1173 & 0.1190 & 0.1153 & \textbf{0.1205} & \textbf{0.1170} & 0.1168 & 0.1166 \\
\hline
\end{tabular}
}
\end{center}
\caption{\label{tab:allfits2j}
Summary of $\alpha_s(M_Z)$ determinations obtained from fits in different mass schemes from~\refcite{Nason:2025qbx}. 
The thrust results, which were the focus of this workshop, are highlighted.}
\end{table}

I.~Stewart presented a field-theoretic framework based on \refcite{Mateu:2012nk}, where hadron-mass effects are systematically incorporated into the structure of leading power corrections. 
The key ingredient is the transverse velocity of the hadron, $r = p_\perp / m_\perp$, which ranges from $r = 0$ for hadrons much heavier than their transverse momentum to $r = 1$ in the massless limit. 

In the massless case, the leading power correction is parametrized by a single constant $\Omega_1$. 
With finite hadron masses, this universal quantity becomes a function of transverse velocity, $\Omega_1(r)$. 
Since thrust is measured inclusively, the power correction for an event shape $e$ involves an integral over all transverse velocities,
\begin{equation}
\Omega_1^{g_e} = c_e \,  \int_0^1 \df r \, g_e(r) \, \Omega_1(r)\,,
\end{equation}
where $c_e$ is a calculable coefficient determined by the rapidity weighting of the observable in the massless limit, $\Omega_1(r)$ is a universal non-perturbative matrix element, and $g_e(r)$ is a weight function that depends on both the observable and the mass scheme. 

Two event shapes belong to the same universality class if they share the same $g_e(r)$, and hence the same non-perturbative parameter $\Omega_1^{g_e}$. 
This has important implications for fits:
\begin{itemize}
    \item For a single event shape, hadron-mass effects can be absorbed into a single parameter $\Omega_1^{g_e}$. Different mass schemes will lead to different fitted values of $\Omega_1^{g_e}$, but within the universality framework they should not affect the extracted $\alpha_s$, provided all data are defined in the same scheme.
    \item For fits combining multiple observables from different universality classes, a single non-perturbative parameter should not be used, even in the dijet limit.
\end{itemize}

Within this framework, the authors of \refscite{Abbate:2010xh,Benitez:2024nav} argue that hadron-mass effects do not represent an additional source of uncertainty in their thrust analyses, for the following reasons. First, in all LEP experiments, thrust is defined purely in terms of three-momenta, so the $p$-scheme should correspond directly to the experimental definition of the observable. 
In this case, theoretical predictions obtained in the same scheme should be suitable for comparison with the data.
Second, their analyses fit a single observable, thrust, in a dijet-enriched region $\tau \in [0.05, 0.15]$, and determine the non-perturbative parameter $\Omega_1^{g_e}$ directly from the data. 
Within the universality framework discussed above, a change of mass scheme should be reabsorbed into a redefinition of $\Omega_1^{g_e}$, thus not affecting the extracted value of $\alpha_s$ at leading power. During the subsequent discussion, it was concluded that a key aspect of this approach is the validation of the above assumptions.

Under the assumptions discussed above, the analyses of \refscite{Abbate:2010xh,Benitez:2024nav} do not assign an additional uncertainty associated with hadron-mass effects beyond the uncertainty on $\Omega_1^{g_e}$. 

\begin{figure}
\begin{center}
    \includegraphics[width=0.7\linewidth, page=1]{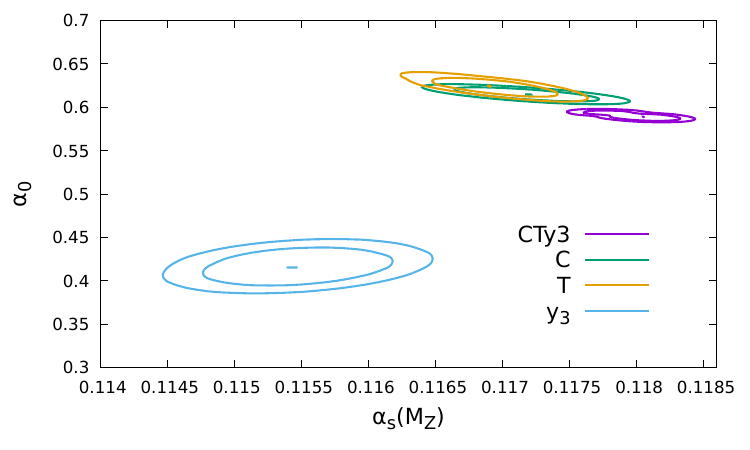}
\caption{Fit contours in the $(\alpha_s, \alpha_0)$ plane for different event shapes. Event shapes in the same universality class show consistent contours, while $y_3$, which belongs to a different class, appears to prefer a different value of the non-perturbative parameter $\alpha_0$.}
\label{fig:universality}
\end{center}
\end{figure}

For fits to multiple observables, such as those performed in \refcite{Nason:2025qbx}, a concern was raised about the assumption of using a single non-perturbative parameter $\alpha_0$ across observables that may belong to different universality classes (with $C$ and $\tau$ on the one hand and $y_{3}$ on the other; see \fig{universality}).

The action items listed below are intended to clarify these issues.

\subsubsection{Discussion outcomes}

The discussion led to several concrete action items and points of agreement:
\begin{itemize}
    \item To test the framework assumptions, the group of \refcite{Benitez:2024nav} will perform fits in the $E$-scheme to verify that changing the mass scheme corresponds primarily to a redefinition of the non-perturbative parameter, without affecting $\alpha_s$. Deviations in single-observable fits due to mass-scheme effects may indicate contamination from corrections beyond the dijet limit.
    \item Experimental groups expressed willingness to provide measurements of event-shape observables in both the $E$- and $p$-schemes. Having data in multiple schemes is highly valuable, as translating between schemes using unfolded data relies less on MC modeling than constructing a transition matrix entirely from simulations.  For each scheme, the four-momentum definitions of selected particles are modified accordingly in both data and MC, and the full analysis chain is repeated. This yields an independent measurement of the observable per scheme. The degree of consistency between these measurements and the result obtained via the MC transfer procedure is then quantified.
    \item Despite its simplicity and resilience to details of the hadronization level (see \refcite{Salam:2001bd}), analyses in the $D$-scheme will not be pursued at this stage, as implementing the $D$-scheme requires replacing all particles, including well-measured charged tracks, with pairs of massless pseudoparticles.
\end{itemize}

\subsection{Impact of resummation space choice}
\label{sec:resum}

A central point of discussion concerned the choice of space in which the resummation of large logarithms is performed. In the presence of multiple soft and/or collinear emissions, the kinematic constraint imposed by the $\tau$ observable factorizes multiplicatively when expressed in Laplace space, with conjugate variable $N$. Consequently, analytic formulations of thrust resummation are generally more straightforward in Laplace space. A similar consideration applies within the SCET framework: while, in momentum (direct) space, the factorization theorem for thrust involves a convolution of the soft function with the jet functions, it reduces to a simple multiplicative structure upon transformation to Laplace space.

Nevertheless, it is also possible to formulate the resummation analytically in direct space by truncating and systematically including subleading towers of contributions order by order in the logarithmic counting, as demonstrated, for example, in the CTTW formalism~\cite{Catani:1992ua}. The equivalence of this approach with the SCET formulation has been established in \refcite{Almeida:2014uva}. In practice, direct-space approaches remain the most commonly used for thrust resummation.

The exact inversion of the Laplace-space resummed expression can only be performed numerically and requires a prescription to handle the Landau pole at large values of $N$. This strategy has been employed recently in \refcite{Aglietti:2025jdj}, supplemented by the adoption of the Minimal Prescription to regulate the Landau pole. In general, results obtained in different resummation spaces are formally equivalent at any finite logarithmic accuracy, differing only by terms beyond the claimed accuracy. However, the numerical impact of such subleading differences has been the subject of recent discussion~\cite{Aglietti:2025jdj}.

\begin{figure}[h!]
\centering
\begin{minipage}[t]{1.1\textwidth}
\vspace{16pt}
\includegraphics[width=.47\linewidth]{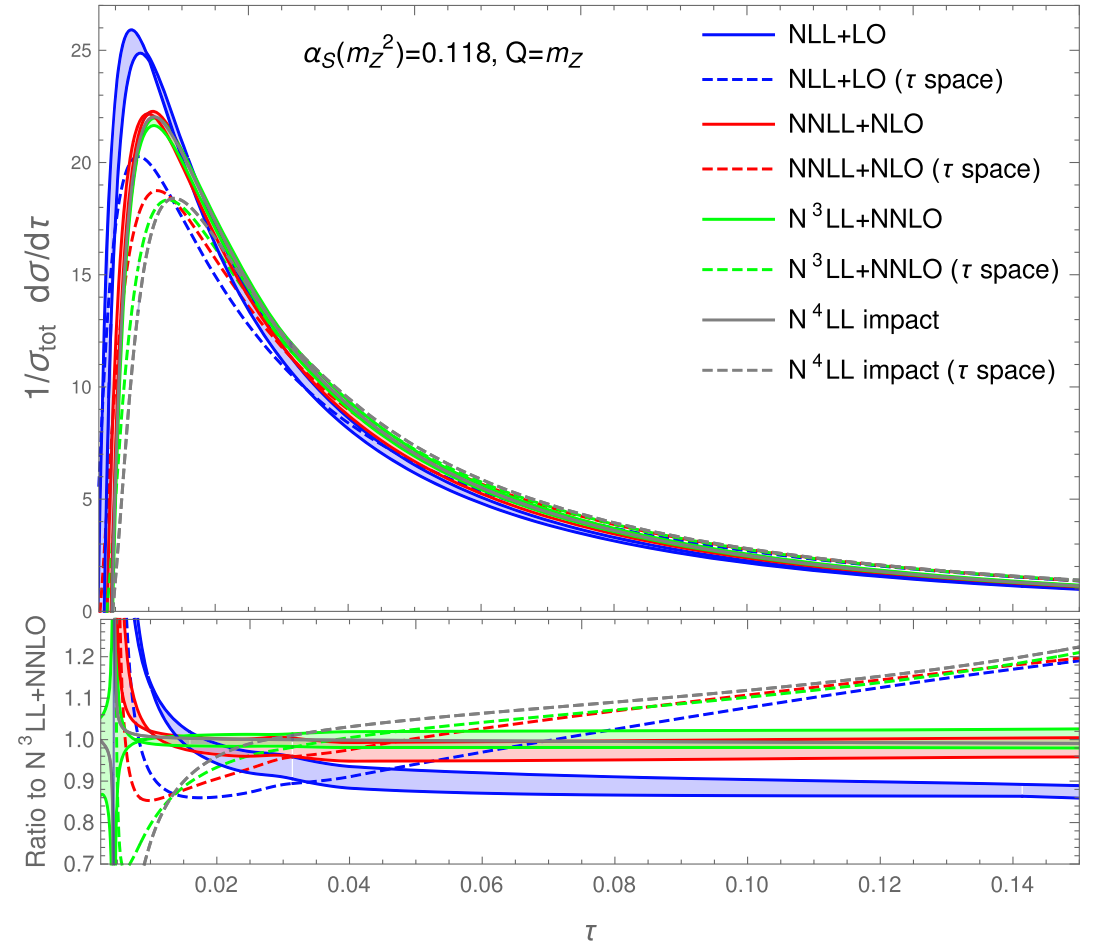}\qquad \includegraphics[width=.4\linewidth]{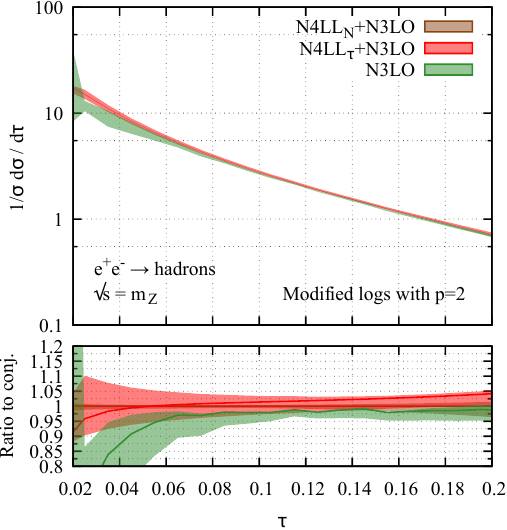}\\[-0.2em]
\hspace*{4.2cm}\textbf{Ref.~\cite{Aglietti:2025jdj}}\hspace*{7.4cm}\textbf{Ref.~\cite{Buonocore:2026kgu}}\\
\end{minipage}
\hspace*{-35pt}
\begin{minipage}[t]{\textwidth}
\vspace{0pt}
\centering
\includegraphics[width=.5\linewidth]{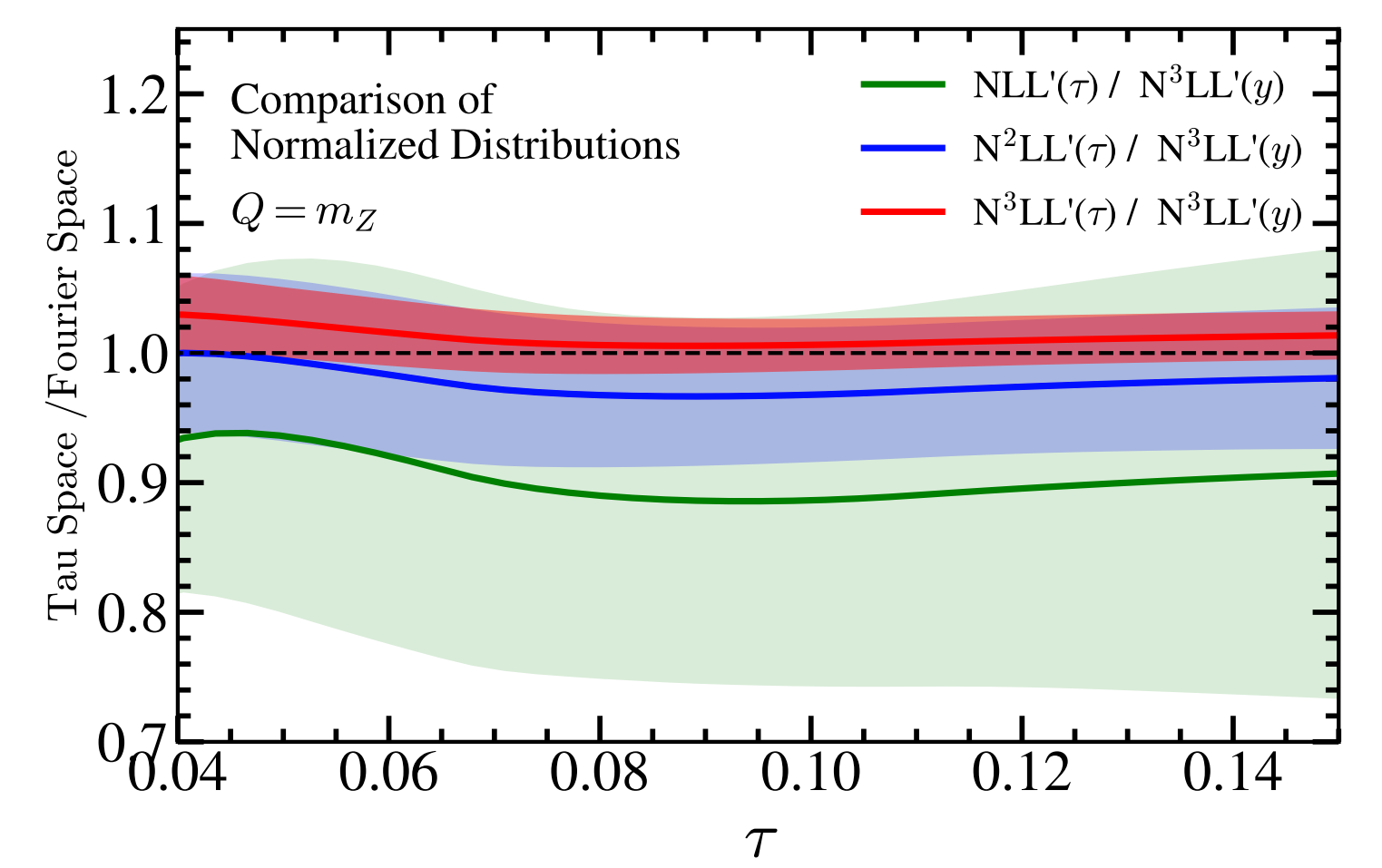}\\[-0.2em]
\phantom{}\hspace{1cm}\textbf{Ref.~\cite{BenitezTalk}}
\end{minipage}
\caption{Comparison of the impact of the choice of resummation space on thrust predictions, as presented by the three groups at the kick-off meeting.
The upper panel shows the two results obtained within the CTTW formalism.
The left plot, from~\refcite{Aglietti:2025jdj}, exhibits sizable differences between direct- and conjugate-space predictions that persist over the moderate- and large-$\tau$ region.
The right plot shows the results of~\refcite{Buonocore:2026kgu}, highlighting substantially smaller discrepancies and compatibility between the direct- and conjugate-space predictions within the estimated perturbative uncertainties in the region to the right of the peak.
Finally, the lower panel displays the SCET results presented at the workshop \cite{BenitezTalk} using the formalism of~\refcite{Benitez:2024nav}, obtained in both direct and conjugate space over the fit region adopted in that analysis, and found to be in agreement with the conclusions of~\refcite{Buonocore:2026kgu}.}
\label{fig:resummation_space}
\end{figure}

During the meeting, G.~Ferrera presented the results of \refcite{Aglietti:2025jdj}. The main findings are shown in the top left panel of \fig{resummation_space}. Significant discrepancies were observed between direct-space and conjugate-space parton-level perturbative predictions for the thrust spectrum. The largest differences are observed in and below the peak region; however, in this regime non-perturbative effects are substantial, making the applicability of perturbative resummation questionable.
More unexpectedly, sizeable deviations persist at moderate values of $\tau$, reaching up to $20\%$ at $\tau = 0.15$. This lies within the conventional fitting range and therefore has clear potential to impact extractions of $\alpha_s$ at the percent level.
Notably, in the analysis of \refcite{Aglietti:2025jdj} these discrepancies do not diminish with increasing perturbative accuracy (from NLL to N$^4$LL) or with higher matching orders. 

In a follow-up contribution, M.~Benitez presented results based on an SCET implementation (bottom panel of \fig{resummation_space}), showing that predictions obtained in direct and Laplace space exhibit a clear pattern of convergence with increasing perturbative order within the fit range $\tau \in [0.05,0.15]$ considered in~\refcite{Benitez:2024nav}. Moreover, the predictions in Laplace-conjugate and $\tau$ space were shown to be consistent within the perturbative uncertainties estimated according to the methodology of~\refcite{Benitez:2024nav}.
Such a theory-error estimate relies on a simultaneous random scan over ten independent theory-variation parameters (such as hard, jet, and soft renormalization scales, as well as transition parameters to assess uncertainties in matching to the fixed-order prediction), and therefore differs substantially from the method used to estimate perturbative uncertainties in resummed event-shape predictions by varying a single renormalization scale, the so-called \emph{3-point variation}, adopted in several works, including \refcite{Aglietti:2025jdj}.

L.~Buonocore then discussed preliminary results from the analysis of \refcite{Buonocore:2026kgu}, focusing on resummation ambiguities associated with the choice of resummation space. 
A key aspect of this study is the behavior of the series corresponding to logarithmically subleading towers of terms that arise from the analytical inversion of the Laplace-space resummation formula. 
The resulting perturbative expansion is characterized by coefficients that grow with the perturbative order, indicating the asymptotic nature of the series. 
In certain contexts, such as threshold resummation, this growth can be factorial, implying sizeable non-perturbative power corrections~\cite{Catani:1996yz}. 
However, this behavior is not observed in the case of thrust: the growth of the coefficients is milder and leads instead to exponentially suppressed contributions, which can nevertheless become numerically relevant in the vicinity of the peak region.
L.~Buonocore presented a preliminary version of the results of \refcite{Buonocore:2026kgu}, obtained with an independent implementation based on the CTTW formalism and including matching to fixed order.
At the highest available accuracy (N$^4$LL matched to N$^3$LO), the residual differences between resummation schemes are found to be around $4\%$ for $\tau \in [0.06, 0.15]$, rather than the $\sim 20\%$ reported in \refcite{Aglietti:2025jdj}, and are encompassed by a sufficiently reliable estimate of perturbative uncertainties, see the top right panel of \fig{resummation_space} where the results of \refcite{Buonocore:2026kgu} are shown. 

The behavior observed in the analyses presented by Benitez and Buonocore is qualitatively consistent with theoretical expectations: increasing perturbative accuracy should progressively reduce the sensitivity to formally subleading choices, such as the resummation space or the matching scheme, provided one remains within a kinematic region where perturbation theory is reliable. 
From this perspective, theoretical uncertainty estimates should be sufficiently conservative to encompass residual differences, including those associated with the choice of resummation space.
More generally, the ability to compare both central values and uncertainty estimates across successive perturbative orders provides a powerful tool on the assessment of perturbative uncertainties and ambiguities affecting theoretical predictions and determinations.

\subsubsection{Discussion outcomes}
As a starting point, the discussion made it clear that differences between results at the same logarithmic accuracy obtained in different resummation spaces or formalisms should be interpreted as ambiguities due to the truncation at a finite order in the perturbative expansion (both in the log-counting sense for the resummation and the traditional fixed order sense for the part related to the matching), rather than as evidence that one resummation space is intrinsically correct and the other is an approximation. 
At finite logarithmic accuracy, different resummation-space choices may differ by subleading terms beyond their nominal accuracy, and the relevant question is therefore the numerical size of these differences and whether they are covered by a reliable estimate of perturbative uncertainties.

The origin of the discrepancies among the different studies regarding the size of these ambiguities was not resolved during the meeting and remains an open issue.
Clarifying the origin of the larger differences between the two resummation formalisms reported in~\refcite{Aglietti:2025jdj} is therefore identified as a high priority. 
Moreover, resummed predictions matched to fixed order should approach the fixed-order result within uncertainties at large $\tau$, irrespective of the resummation space, which is not observed in the results of \refcite{Aglietti:2025jdj} shown in \fig{resummation_space}.
This issue has a direct impact on the extraction of $\alpha_s$ and on the corresponding uncertainty estimates. The following concrete action items were identified:

\begin{itemize}
\item Perform an ingredient-by-ingredient comparison within the CTTW formalism, where the two implementations are closest in their theoretical organization but nevertheless lead to substantially different numerical conclusions.
The comparison should first be carried out at the purely resummed level, before introducing matching to fixed order.
This provides the cleanest setup in which to compare the perturbative radiator, logarithmic counting, treatment of subleading towers, and prescription for the inverse transform.
Once the origin of any discrepancy at the resummed level is understood, the benchmark should be extended step by step to include matching to fixed order.
We note that matching to fixed order should not increase the discrepancy. On the contrary, by fixing subleading terms to the unambiguous fixed-order prediction, it should further reduce the differences between the two formulations.

\item Perform a more systematic assessment of matching uncertainties.
In particular, the impact of the matching procedure should be studied order by order, both at the level of the thrust distribution and of the extracted value of $\alpha_s$.
Concrete tests include comparing different matching schemes and varying the rate at which the resummation is turned off during matching, for example by modifying the power and transition point in the modified-log prescription~\cite{Bozzi:2005wk,Monni:2011gb,Gehrmann:2012sc}.
Profile-function-based matching prescriptions, such as those employed in the SCET fits, contain analogous parameters that control the perturbative ambiguities associated with the matching procedure.
These parameters are varied as part of the theory uncertainty estimate in SCET analyses, and it would be valuable to quantify their individual impact on the extracted value of $\alpha_s$ in the fit of \refcite{Benitez:2024nav}, as it was done in figure~16 of \refcite{Abbate:2010xh}.
\end{itemize}

\section{Conclusions, Recommendations, and Outlook}
\label{sec:conclusions}

The kick-off meeting successfully brought together key experts from both the experimental and theory communities and, while focusing on thrust, enabled a productive discussion of important open issues in event-shape determinations of $\alpha_s$.
It confirmed the value of a dedicated task-force format for addressing questions that require sustained technical exchange between the groups involved.
The meeting identified a concrete programme of tests, comparisons, and experimental inputs needed to determine which assumptions are justified and where additional uncertainties or theoretical ingredients are required.
The recommendations below summarize the immediate work programme that emerged from the discussions.

\subsection{Recommended actions for experimental groups}

The experimental reanalysis of archived LEP data is a central component of the task-force programme.
The new ALEPH and DELPHI thrust results already indicate shifts with respect to the legacy publications that are large enough to affect modern high-precision fits, and understanding their origin is therefore a first priority.

\begin{enumerate}
\item \textbf{Clarify the origin of the shifts with respect to legacy data.}
The ALEPH and DELPHI reanalyses should identify which elements of the updated analyses are responsible for the observed shifts in the thrust distribution.
This includes separating effects that reflect clear methodological improvements from effects associated with different, but defensible, analysis choices.
Where the latter occur, their impact should be documented as a potential source of experimental systematic uncertainty for future fits.

\item \textbf{Provide the experimental information needed for precision fits.}
Reanalyses should provide detailed documentation of the unfolding procedure, systematic variations, detector limitations, the impact of the MC corrections, and the treatment of effects such as neutral-particle response and initial-state radiation.
Whenever possible, covariance information should be provided across bins, center-of-mass energies, and observables, since this is essential for combined fits and for assessing correlations between theoretical and experimental systematics.

\item \textbf{Extend the prioritized set of observables and energies.}
The immediate priority is thrust, but the next set of measurements should include observables that are central to existing and future $\alpha_s$ studies, in particular the $C$-parameter and $y_3$.
Measurements at LEP~1 and LEP~2 energies, and where possible lower-energy data, provide an important lever arm for testing the $Q$ dependence of perturbative and non-perturbative effects.
Two-particle correlation measurements should also be extended beyond energy correlations on tracks, including charge correlations and energy-energy correlations with neutral particles.

\item \textbf{Provide measurements in multiple mass schemes.}
Dedicated measurements in both the $p$- and $E$-schemes should be pursued for the relevant event-shape observables.
These measurements are needed to test whether mass-scheme changes in single-observable dijet fits are absorbed by the fitted non-perturbative parameter, or whether residual shifts in $\alpha_s$ indicate contamination from beyond-dijet effects.
They also reduce the reliance on MC-only transformations between schemes and provide inputs directly matched to the theoretical assumptions being tested.

\item \textbf{Provide projections for future $e^+e^-$ collider measurements and assess detector requirements (if needed).}
The expected precision in $e^+ e^- $ measurements at the FCC-ee will be fully dominated by systematic rather than statistical uncertainties, given the very large available data samples. 
Dedicated projections should be developed to quantify the achievable precision of event-shape measurements and related $\alpha_s$  determinations at FCC-ee energies. 
These studies should identify the leading experimental systematic uncertainties, including detector response, particle reconstruction, calibration, unfolding, and acceptance effects, among others. 
The impact of these uncertainties on the ultimate physics reach should be evaluated, together with the implications for FCC-ee detector requirements, reconstruction strategies, and analysis design. 
This will help establish the detector performance targets needed to fully exploit the precision QCD potential of future  $e^+ e^-$ data.
\end{enumerate}

\subsection{Recommended actions for theory groups}

The theory recommendations are organized around controlled tests of assumptions that currently differ across analyses.
The purpose is not to select a preferred framework a priori, but to identify which assumptions are valid in which regimes and to define benchmarks that can be applied consistently across observables and processes.

\begin{enumerate}
\item \textbf{Study of the size of effects beyond the dijet factorization.}
The separation between dijet and 3-jet regimes should first be clarified at the perturbative level.
A possible source of useful information is the order-by-order comparison of the non-singular partonic cross section, i.e. the perturbative part of the fixed order cross section which is not described by the leading power factorization theorem, with the full fixed order cross section, in particular testing whether the observed size of the non-singular contribution in the fit region remains a small part of the full correction to the partonic cross section at each perturbative order. This exercise should not be limited to thrust only and it would be valuable to have it carried out also for other observables where resummation is commonly applied and higher order resummation and fixed order information is available.

\item \textbf{Perform fit in the transition between resummed and fixed-order descriptions.}
Thrust fits in the transition region, for example N$^3$LL$^\prime+\mathcal{O}(\alpha_s^3)$ fits in $\tau \in [0.15,0.33]$, should be compared with dijet fits in $\tau \in [0.06,0.15]$ within the same theoretical framework.
This comparison is needed to assess whether the treatment of the transition from resummation to fixed order leads to compatible determinations of $\alpha_s$.

\item \textbf{Perform controlled hadron-mass and mass-scheme tests.}
Single-observable thrust fits should be repeated in both the $p$- and $E$-schemes.
In a fit genuinely restricted to the dijet limit, the leading hadron-mass dependence should be absorbed into the fitted non-perturbative parameter; a significant residual shift in $\alpha_s$ may diagnose contamination from beyond-dijet effects, including 3-jet power corrections, or the presence of other effects not accounted for in the theoretical description.

\item \textbf{Resolve the reported large resummation-space effect through targeted benchmarks.}
  A sizeable dependence on the choice of resummation space has been reported in the calculation of \refcite{Aglietti:2025jdj}, while two independent analyses presented at the workshop found different results, i.e., smaller effects in the $\tau \in [0.06,0.15]$ region and the expected reduction of the ambiguity with increasing perturbative accuracy.
  This tension should be resolved through a targeted benchmark comparison, starting from the setup in which the large effect was observed and then aligning, step by step, the
  perturbative ingredients, matching conventions, fit ranges, prescriptions for the inverse transform, and uncertainty variations.
  The first target should be the thrust fit region $\tau \in [0.06,0.15]$, followed by an extension to broader ranges.
  The residual uncertainties induced by resummation spaces should be encompassed by the estimate of the perturbative theoretical uncertainties, and can therefore serve as a diagnostic of the reliability of such an estimate.
  At present, however, the task force does not recommend assigning a separate additional theory uncertainty for the large $15-20\%$ effect reported in \refcite{Aglietti:2025jdj}, which has not been confirmed by \refcite{Buonocore:2026kgu} or by the independent SCET implementation presented by M. Benitez and whose origin remains to be understood.
  Once the benchmark comparisons have led to a convergent assessment of the size and origin of the effect, its treatment in the uncertainty budget, including whether an additional independent uncertainty is warranted, should be discussed in future task-force meetings.
\item \textbf{Assess matching and profile uncertainties in $\alpha_s$ fits.}
Matching uncertainties should be studied both at the level of the matched thrust distribution and at the level of the extracted value of $\alpha_s$.
Different matching schemes should be compared directly using common perturbative inputs and fit settings, so that the size of formally subleading terms introduced by the matching prescription can be assessed separately from scale variations and from the resummation-space ambiguity.
The rate with which resummation is turned off in matching should be varied systematically, with the impact assessed order by order on both the matched spectrum and the fitted value of $\alpha_s$, since this probes the stability of the transition to the fixed-order description in the upper part of the fit range.
For SCET fits, the impact of profile-function parameters on the extracted $\alpha_s$ should be broken down in a form analogous to figure~16 of \refcite{Abbate:2010xh}.

\end{enumerate}

\subsection{Coordination and longer-term outlook}
\label{sec:outlook}

The task force should maintain a documented set of benchmark inputs, plots, fit settings, and follow-up results so that progress can be tracked transparently between meetings.
The TWiki page of the task force~\cite{twiki} will provide a common entry point for collecting materials, documenting follow-up exercises, and maintaining the evolving list of benchmarks and priorities.
Each comparison should specify the observable definition, data input, perturbative accuracy, matching prescription, fit range, non-perturbative parametrization, and uncertainty procedure used.
This documentation is essential for turning the outcome of the task force into a durable community reference for future high-precision determinations of $\alpha_s$.

Future meetings should be organized around the concrete follow-up exercises identified in this document, with the goal of reviewing completed cross-checks, identifying remaining discrepancies, and updating the list of priorities.
While the kick-off meeting focused on thrust, subsequent activities should broaden the scope to additional event shapes and correlators, multi-observable fits, theory correlations, and other precision QCD observables at lepton colliders according to the needs and requests of the community.
Several topics that were not discussed in detail at the kick-off meeting are nevertheless central to the broader task-force programme.

Fits in the peak region are an important example.
In this region non-perturbative effects are large. 
The hierarchy $\Lambda_{\rm QCD} \ll Q\tau$ no longer holds, removing the parametric control over further non-perturbative parameters, such as higher moments of the non-perturbative shape function. 
Experimental resolution effects are also especially delicate.
Precision fits in this region therefore require a dedicated treatment rather than a direct extrapolation of the tail-region assumptions.

Similarly, renormalon-subtraction frameworks and the definitions of the corresponding non-perturbative parameters remain important elements of precision event-shape fits~\cite{Hoang:2007vb,Abbate:2010xh}.
Such frameworks can employ different scheme choices for the renormalon subtraction and the assessment of their residual perturbative scheme dependence is part of the systematic uncertainty studies in high-precision strong-coupling extractions~\cite{Bell:2023dqs,Benitez:2024nav}. 
These scheme choices are of interest for further comparisons, including in relation to the ambiguity associated with leaving the leading renormalon unsubtracted.

Non-perturbative power corrections also raise broader theoretical questions beyond the specific dijet-versus-3-jet discussion of the kick-off meeting.
The anomalous scaling of linear power corrections and hadron mass effects is a particularly interesting direction~\cite{Salam:2001bd,Mateu:2012nk,Dasgupta:2024znl,Chen:2024nyc,Farren-Colloty:2025amh,Banfi:2025crj,Chen:2026hmd}.
It will become increasingly relevant for event-shapes that are studied across a wide range of center-of-mass energies, including the low-energy QCD programme proposed for FCC-ee~\cite{dEnterria:2025pml}.

Another major theme for future work is the treatment of theory correlations and perturbative uncertainties beyond conventional scale variations, including theory nuisance parameter approaches~\cite{Tackmann:2024kci} and other methods designed to encode correlated theory systematics in single- and multi-observable fits.
Further developments in next-to-leading-power resummation should also be incorporated into the task-force discussions, both as a way of improving predictions and as a diagnostic for the domain of validity of leading-power descriptions~\cite{Moult:2018jjd,Moult:2019uhz,Beneke:2022obx}.

Several additional ingredients involve both theory and experiment.
QED effects, initial-state radiation, and the use of lepton parton distribution functions are essential for a fully controlled comparison between theory predictions and high-precision lepton-collider data~\cite{Frixione:2019lga,Bertone:2019hks,Bertone:2022ktl,Dittmaier:2026nae}.
The task force should also keep track of the rapidly expanding field of weighted cross sections and collider correlators, which provides a complementary view of perturbative radiation, hadronization, flavor and charge flow, and the transition between partonic and hadronic descriptions~\cite{Basham:1978bw,DelDuca:2016ily,Tulipant:2017ybb,Kardos:2018kqj,Moult:2025nhu,Chen:2023zlx,Lee:2024esz,CMS:2024mlf,Lee:2025okn,Chang:2025kgq,Chen:2026lsc,Chen:2020vvp,Jaarsma:2025tck,Electron-PositronAlliance:2025fhk,Monni:2025zyv}.
Examples include energy-energy correlations and projected energy correlators, as well as track- and charge-weighted variants that connect naturally to experimentally robust measurements, a connection already visible at the kick-off meeting in the track energy-energy correlator discussed in the experimental session.
For precision applications, these observables require dedicated theoretical inputs, including resummation, non-perturbative corrections, track functions, moments of track functions, and their evolution~\cite{Chang:2013rca,Chang:2013iba,Chen:2020vvp,Li:2021zcf,Jaarsma:2023ell}. The observables and topics discussed here are moreover directly connected to transverse-momentum-dependent (TMD) physics, and pursuing this connection at LEP energies would yield information complementary to ongoing Belle II measurements~\cite{Belle:2019ywy, Fernandez:2026nya} while informing the future HL-LHC, FCC, and Electron-Ion Collider (EIC) programs~\cite{Boussarie:2023izj}.

Alongside this expanding correlator programme, extending high-precision studies across the canonical event shapes would recover a strength of the LEP programme: comparisons among thrust, the $C$-parameter, heavy jet mass, total and wide jet broadenings, and $y_3$ can expose observable-dependent limitations of perturbative and non-perturbative descriptions that are hidden in any single distribution~\cite{ALEPH:2003obs,OPAL:2004wof}.
Recent efforts to extend state-of-the-art precision to heavy jet mass combine high-order dijet predictions with the resummation of its Sudakov shoulder~\cite{Benitez:2025vsp,Bhattacharya:2023qet}.
The corresponding extraction employs two independent non-perturbative parameters to describe the dijet and three-jet regions separately, with the latter accounting for the large negative three-jet power correction.
At the same time, several aspects of how these additional perturbative and non-perturbative ingredients should be organized and assessed remains a subject for further discussion.
This lesson is also evident in the analyses of \refscite{Nason:2023asn,Nason:2025qbx}, where three-jet power corrections were studied for a broad set of shapes and found to have markedly different behavior, including negative and rapidly varying corrections for the hemisphere-mass observables.
Taken together, these developments motivate efforts to bring as many observables as possible to the highest perturbative accuracy achievable for each, together with a controlled description of their non-perturbative corrections.

Turning this broader set of predictions into precision extractions brings into sharper focus several questions that are already central in single-observable fits: the treatment of theory correlations, robust uncertainty estimates, and well-motivated criteria for determining the number of independent non-perturbative parameters.
These issues remain nontrivial even when one observable is analysed in a controlled fit region, while joint fits add a further layer of subtlety because the same ingredients must be treated consistently across a larger set of observables, universality classes, and fit regions.
The questions become particularly important when the combined predictions are available at different perturbative accuracies or their non-perturbative corrections have different levels of theoretical control.
The value of a broad programme lies both in its potential precision gains and, crucially, in using comparisons among observables to test QCD in several independent settings.
Interpreting such gains will therefore depend on parallel progress in the perturbative and non-perturbative description of each observable and in the understanding of correlated systematic effects.
Developing a reliable framework for such joint analyses is an important longer-term objective, complementary to resolving the remaining issues for the most precise individual probes. 
The absence of these and the other aforementioned topics from the first meeting reflects the deliberately restricted scope chosen to enable concrete progress.

The task force should also remain open to new technologies that can enhance this research. One example is AI agentic workflows for data preservation, reinterpretation, and theory--experiment comparisons~\cite{Badea:2026klb,Moreno:2026mqk}, provided they are used in a transparent and reproducible manner.
A second example is the opportunity to develop synergies with the lattice-QCD and quantum-simulation communities, whose progress in accessing non-perturbative gauge dynamics could eventually open new ways to constrain matrix elements such as $\Omega_1$ and various moments of the associated non-perturbative shape functions, and to obtain differential information on the latter.
Recent lattice efforts to access light-cone dynamics, from quasi-PDF methods to determinations of the Collins--Soper kernel~\cite{Ji:2020ect,Ebert:2018gzl,LatticePartonLPC:2023pdv,Avkhadiev:2024mgd,Tan:2025ofx}, and quantum-simulation studies of real-time gauge dynamics and string breaking~\cite{Bauer:2022hpo,Banerjee:2012pg,Gonzalez-Cuadra:2024xul,Crippa:2024hso,Ciavarella:2024fzw,Echevarria:2020wct} illustrate the potential for such connections, especially for the Minkowskian matrix elements with multiple lightlike Wilson lines relevant here, which are not directly accessible with standard Euclidean lattice methods.

In the longer term, the task force aims to provide a structured assessment framework for collider determinations of $\alpha_s$, analogous to community-based efforts in lattice QCD. 
Such a framework should not replace individual analyses, but should clarify which assumptions have been tested, which benchmarks have been passed, and which sources of uncertainty remain unresolved.

\section*{Acknowledgements}

We thank the FCC-ee Physics Studies Coordination and the LPCC for supporting this task force and providing the framework for this community effort. 
We are grateful to the CERN Theory Department for hosting the workshop and to the TH workshop secretariat for organizational support. 
We thank all speakers for their presentations and all participants for the stimulating discussions. 
A. Badea is supported by the Schmidt Sciences Foundation. 
I.W. Stewart is supported by the U.S. Department of Energy, Office of Science, Office of Nuclear Physics grant number DE-SC0011090. 
Z. Capatti is supported by the Swiss National Science Foundation grant PCEFP2\_203335.
S. Jaskiewicz is supported by the Swiss National Science Foundation Ambizione grant PZ00P2\_223524.
Y.J. Lee is supported by the U.S. Department of Energy, Office of Science, under Grant No. DE-SC0011088.
\bibliographystyle{jhep}
\bibliography{references}

\end{document}